\documentclass[11pt,twoside,a4paper]{article}
\usepackage[dvips]{graphicx}
\usepackage[english]{babel}
\usepackage{amssymb}
\usepackage{float}
\usepackage{color}
\begin{document}

\title{Pulsar electrons detection in AMS-02 experiment.
       Model status and discovery potential.}
\author{Jonathan Pochon\footnote{email: jgpoochon@iac.es} \\
         \small{\it Instituto de Astrof\'isica de Canarias, C. V\'ia L\'actea, 38200, La Laguna, Tenerife, Spain}}

  \maketitle
\vspace*{-7.cm}
\begin{flushright}
\end{flushright}
\vspace*{7.cm}

  \abstract{The measurements of electrons ($e^{\pm}$) from cosmic rays have begun a new era a
            few years ago with high precision experiments like PAMELA {\bf \cite{exp_pamela}} and
	    Fermi-LAT {\bf \cite{exp_fermi}}. The positron
	    fraction seems to indicate an unknown component above the standard background described
	    in the last 40 years, mostly by HEAT {\bf \cite{heat_exp}}. In the last few years, the
	    PAMELA satellite has confirmed the positron fraction excess above
	    10 GeV, and studying Fermi-LAT data, the electron flux seems to be steeper than expected. While
	    these new measurements have not closed the debate, results from AMS-02 {\bf \cite{ams_exp}}
	    are expected to reach the accuracy needed to determine a full description of this excess
	    and possibly give some evidence on the possible source. We will present in this note, the
	    AMS-02 capacity in the case of positrons produced by pulsars.}

  \section*{Introduction}
  
\begin{figure}[ht!]
     \begin{minipage}[c]{.45\linewidth}
       \centering
       \includegraphics*[height=7.5cm,width=7.5cm]{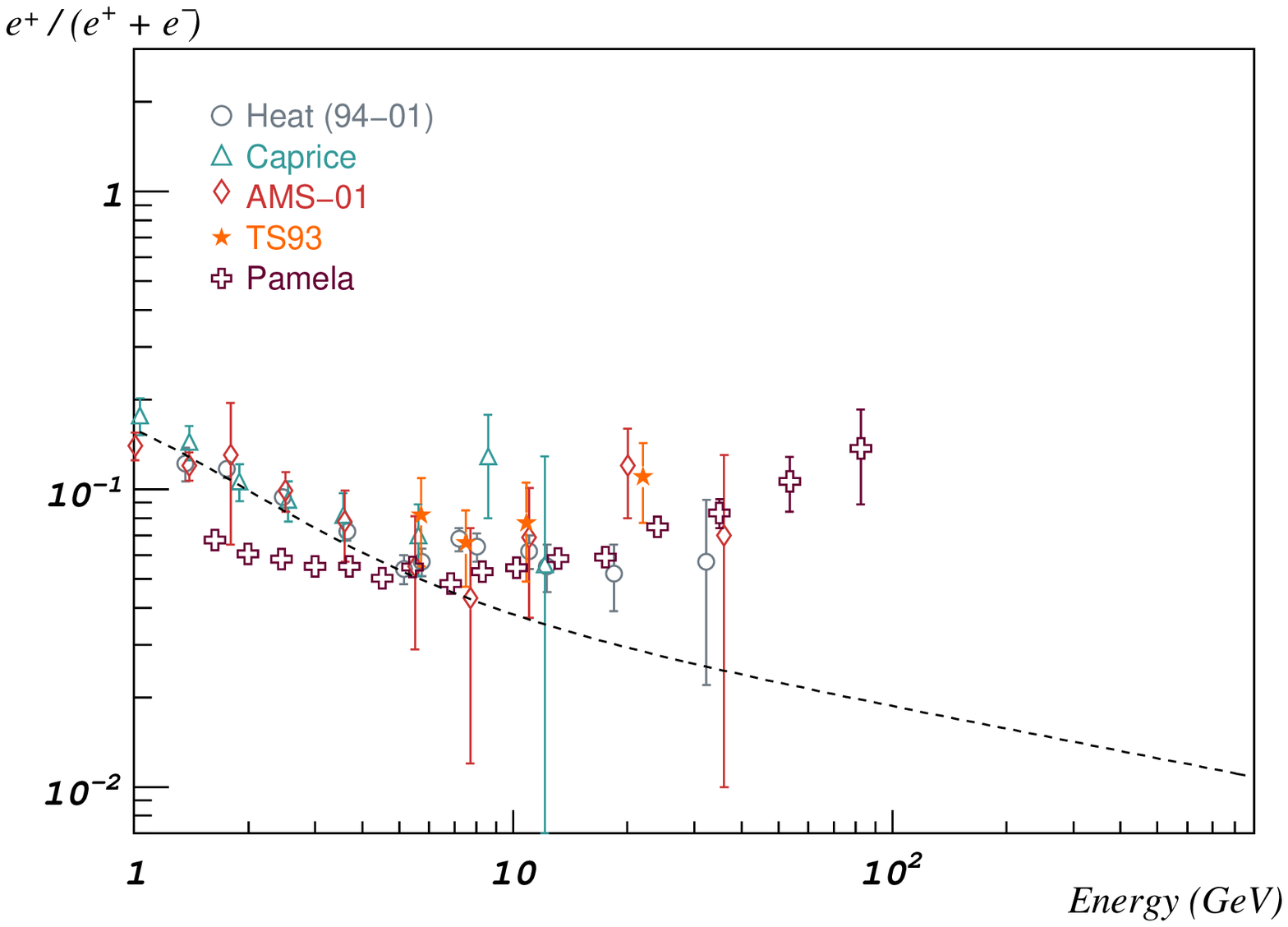}
     \end{minipage}\hspace{0.8cm}
     \begin{minipage}[c]{.45\linewidth}
       \centering
       \includegraphics*[height=7.5cm,width=7.5cm]{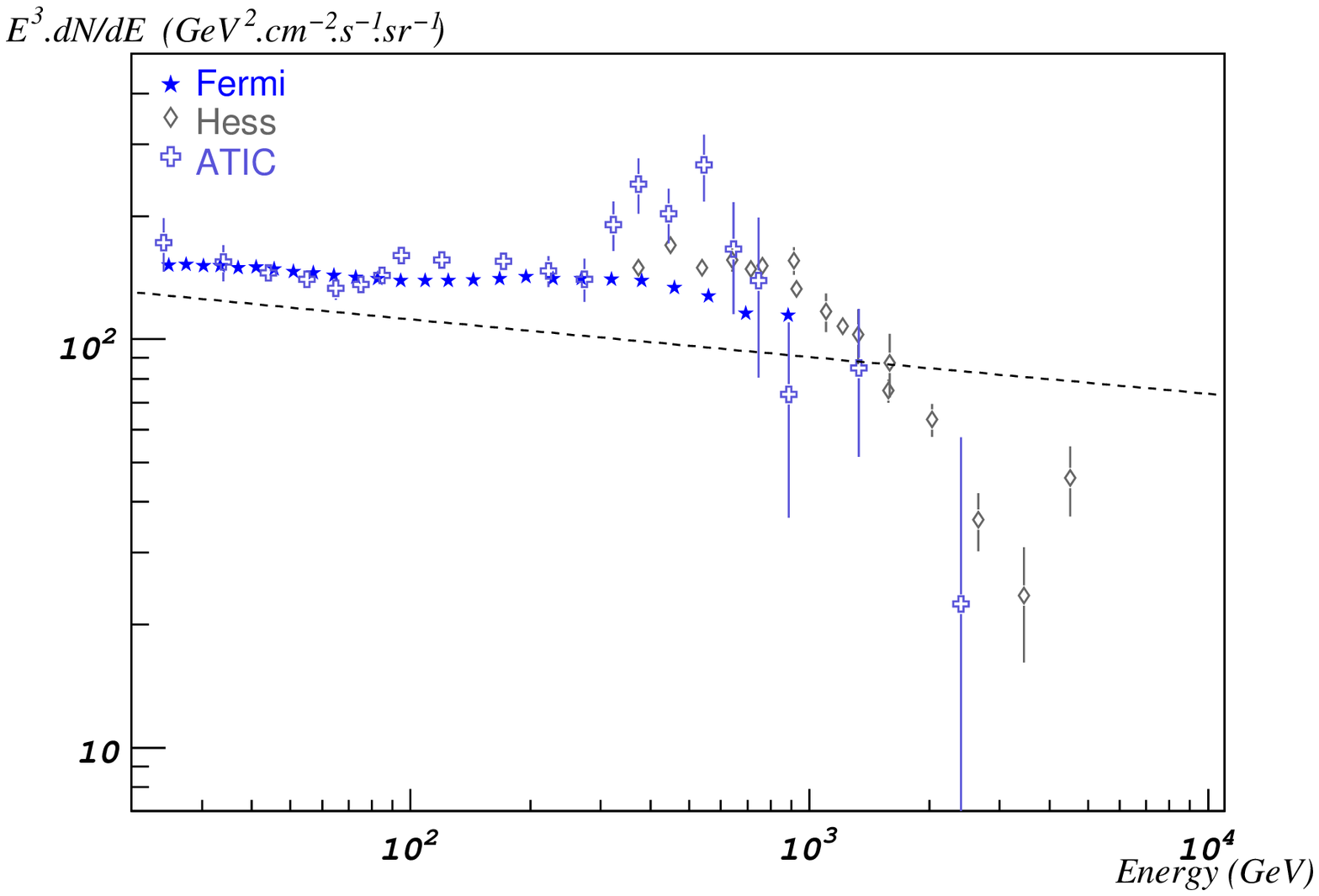}
     \end{minipage}
     \caption{\it Left: Positron fraction for TS93 {\bf \cite{ts93}}, HEAT {\bf \cite{heat}} {\bf \cite{heat2}},
                    Caprice {\bf \cite{caprice}}, AMS-01 {\bf \cite{ams01}} and PAMELA
		    {\bf \cite{dat_pamela}} compared with expected background {\bf \cite{moskalenko}}.
		  Right: Electrons flux for Fermi-LAT, HESS and ATIC compared to the standard background.}
     \label{etat_lieux}
    \end{figure}

    The context of the positron-electron cosmic rays can be summarized by a series of measurements in 
    disagreement with what we understand. Indeed, since the 70s most of the experiments, namely TS93
    {\bf \cite{ts93}}, HEAT {\bf \cite{heat}} {\bf \cite{heat2}}, Caprice {\bf \cite{caprice}}, AMS-01 {\bf \cite{ams01}},
    PAMELA {\bf \cite{dat_pamela}}, exhibit a deviation from the standard framework in the positron
    fraction ($\frac{e^+}{e^+ + e^-}$), represented by figure {\bf \ref{etat_lieux}} (left). Looking at
    the total flux of $e^{\pm}$, observations by  Fermi-LAT  {\bf \cite{dat_fermi}},
    balloon experiments like ATIC {\bf \cite{ATIC}} and ground ones like HESS {\bf \cite{HESS}} present
    a steeper behavior summarized in figure {\bf \ref{etat_lieux}} (right). 
    The standard propagation model can well explain the cosmic rays energy spectrum above 1 GeV, shown in
    studies by Moskalenko and Strong {\bf \cite{moskalenko}}. However, the positron fraction above 10 GeV
    remains unexplained.All models from propagation to source candidates were improved trying to reproduce
    observational data. During propagation, most of the models are consistent between them, the data uncertainty
    allowing high degeneracy models {\bf \cite{maurin}}, in particular for the measurement of the nuclei ratio
    $B/C$, $Be^{9}/Be^{10}$. Regarding source candidates, two approaches seem to be preferred. The first one, is
    to consider the closest astrophysical objects able to produce positron-electron pairs, like e.g. pulsars
    {\bf \cite{taylor_puls}} {\bf \cite{backer_puls}}. Some pulsars are sufficiently close and energetic to be
    responsible of positrons deviation. The second possibility, is to consider dark matter particles, which can
    produce $e^{\pm}$ through annihilation {\bf \cite{cirelli}} {\bf \cite{bergstrom}} or decaying {\bf \cite{zhang_decay}}.
    So far, only PAMELA and ATIC have claimed or confirmed such a deviation from standard background. For
    PAMELA (figure {\bf \ref{etat_lieux}} left) the positron fraction above 10 GeV is growing, while for ATIC,
    the electron flux forms a "bump" around 400 GeV (figure {\bf \ref{etat_lieux}} right). However, the Fermi-LAT
    measurements do not agree with the ATIC bump result, establishing only a mildly harder $e^{\pm}$  flux
    {\bf \cite{dario}}. In this context, AMS-02 will be a powerful detector able to give crucial informations on
    positrons that are directly connected to this possible extra contribution with respect to expected background.
    This note will discuss the topic of electrons produced by pulsars, and will present the AMS-02 capacity.

  \section{Electrons coming from pulsars}
  
     Since the discovery of pulsars in 1967 {\bf \cite{hewish_bell}}, most of the models agree with
     measurements like periodicity, but some unknowns still remain, e.g. the gamma-ray production.
     Electron production in the pulsar has not yet been observed, and their detection would be an
     interesting discovery. Through systematic surveys, pulsars are well studied and basic characteristic
     are available. Figure {\bf \ref{age_dist}} (left) shows that for the Australia Telescope National
     Facility (ATNF) {\bf \cite{ATNF_cata}}, observed pulsars in the Galaxy cover a good range in distance. 
    
    \subsection{Pulsar $e^{\pm}$ production}
     
\begin{figure}[ht!]
      \begin{minipage}[c]{.45\linewidth}
       \centering
       \includegraphics*[height=7.5cm,width=7.5cm]{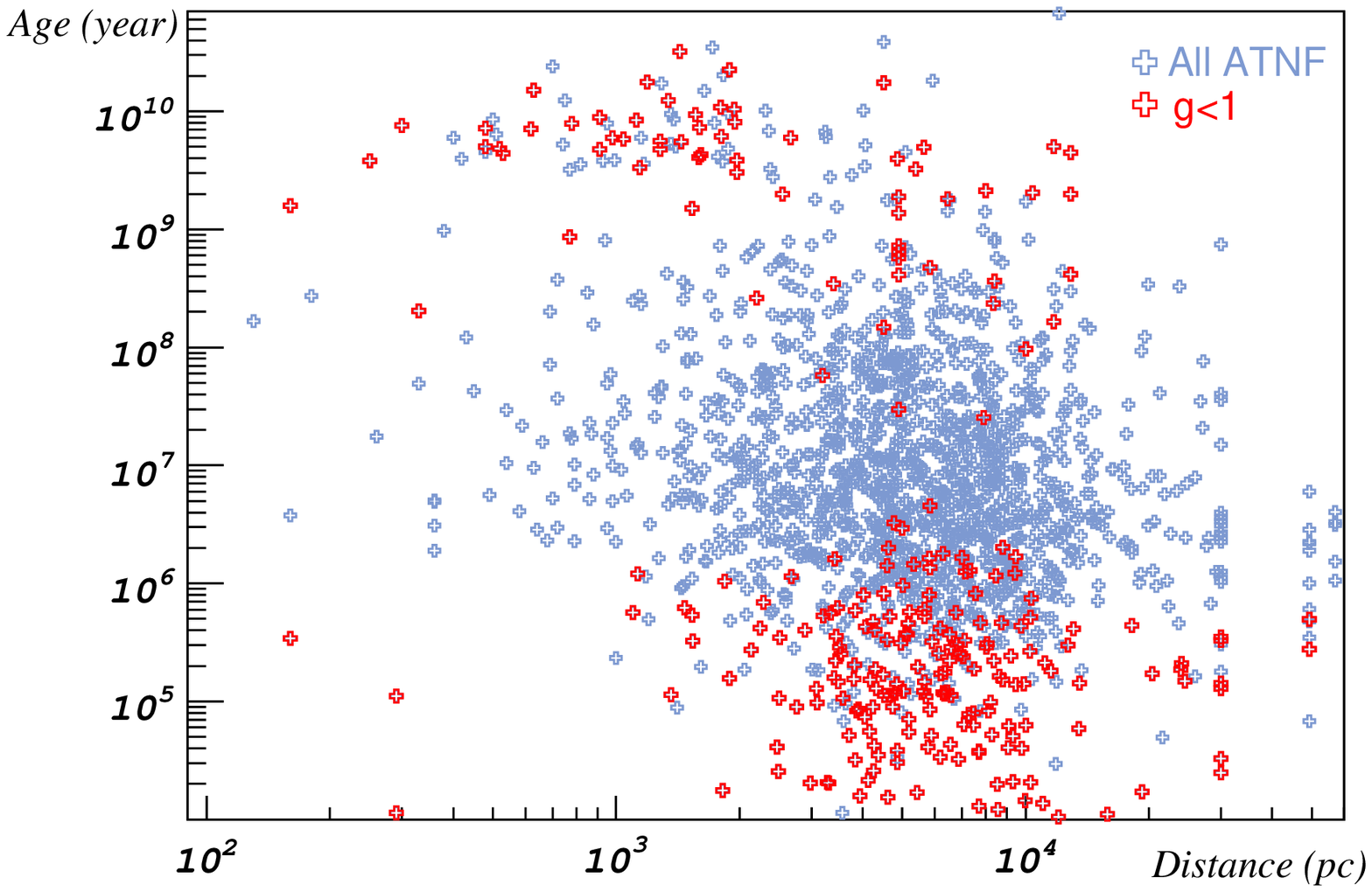}
      \end{minipage}\hspace{0.8cm}
      \begin{minipage}[c]{.45\linewidth}
       \centering
       \includegraphics*[height=7.5cm,width=7.5cm]{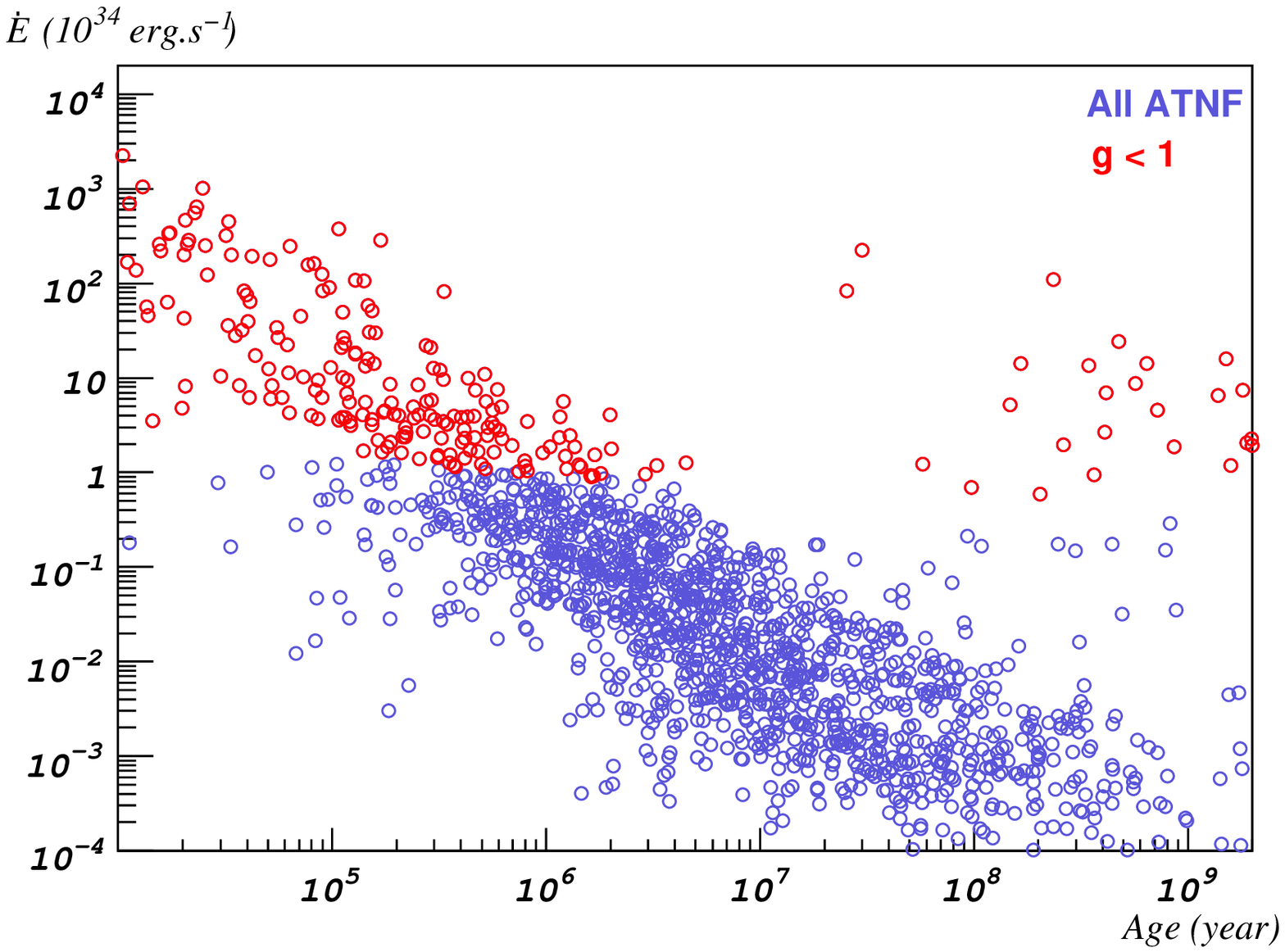}
      \end{minipage}
      \caption{\it Distribution of pulsars from the ATNF catalogue {\bf \cite{ATNF_cata}}
                  including "gamma-ray" ones ($g<1$). Left: Pulsar age versus distance.
		  Right: Puslar spin-down power versus age.}
      \label{age_dist}
     \end{figure}

     Gamma-ray production in pulsars, giving electron-positron pairs, can be modeled by two mechanisms
     of relativistic particle acceleration: the {\it polar cap} (PC) {\bf \cite{rudermans} \cite{arons}} and
     the {\it outer gap} (OG) {\bf \cite{cheng}}. EGRET {\bf \cite{thom_egret}} {\bf \cite{fierro_egret}} has shown evidence,
     for several pulsars, of a pulsed gamma-ray emission at GeV scale, giving confidence to the fact that the
     magnetosphere must be involved for the charged particle acceleration. P. Goldreich and W.H. Julian
     {\bf \cite{julian}} have demonstrated that magnetic field tears away electrons from the pulsar surface.
     The distribution of charged particles in the magnetosphere screens the electric field $E_{||}$ parallel
     to the magnetic field ($\textrm{\bf{B$\cdot$E}} = 0$), allowing the co-rotation of the whole system.
     But in some locations, the condition $\textrm{\bf{B$\cdot$E}} = 0$ is not maintained and particle can be
     accelerated following field lines. From there, a volume named {\it light cylinder} can be defined as a
     cylinder with the rotation axis as symmetry axis, and with radius equal to the co-rotating part. Outside
     this frontier, magnetic field lines are not closed. The acceleration can take place mostly in two locations.
     The first one is close to the surface near the magnetic pole, a situation described by the {\it polar cap}
     model (PC). The other, located between the last closed magnetic field line and the surface along the null
     electric surface defined by the condition ${\bf \Omega\cdot\textrm{\bf B}} = 0$, where ${\bf \Omega}$ is
     the pulsar rotational velocity, a configuration dubbed the {\it outer gap} model (OG). Briefly, in both of
     them, quasi-static electric field accelerates the  relativistic electrons, producing electron-positron pairs.
     The PC model consists of two steps. First one inside the acceleration region, a charged particle ($e^{\pm}$)
     is taken away from the surface due to the electric field radiating gamma-ray by synchrotron. Then these
     gamma-rays create an $e^{\pm}$ pair, through magnetic field or by interaction with thermal X-rays from
     the pulsar surface. These secondaries are moving toward the light cylinder and then can escape. On the 
     other hand, the OG model proposes a pair production from photon-photon interaction. $e^{\pm}$ follow field
     lines radiating gamma-rays which interact with low energy photons (X-rays, infrared) producing  $e^{\pm}$
     pairs, and then able to escape outside the light cylinder. First difference between the two models is that
     the OG location is farther than the PC one, so more distant from the magnetic field giving harder flux,
     while the PC configuration has a bigger contribution at low energy. Chi et al. {\bf \cite{chi_1996}}
     estimate that these two models give comparable total energy output in $e^{\pm}$. A criterion for the OG
     existence is given by Zhang et al. {\bf \cite{zhang}} and implies that $g$, ratio between dimension of the
     OG and radius of the light-cylinder, should be less than one. The OG existence condition can be expressed
     as {\bf \cite{zhang}} 

     \begin{equation}
        g = 5.5 P^{26/51} B_{12}^{-4/7} < 1.
     \end{equation}
    
     with $P$ being the pulsar period in $s$ and $B_{12}$ the pulsar magnetic field in $10^{12}$ G. Conventionally,
     pulsars with the condition $g<1$ are gamma-ray pulsars. The ATNF {\bf \cite{ATNF_cata}} compiled the most
     complete and updated pulsar catalogue{\bf \cite{ATNF_publi}}, comprising 1794 pulsars, with 272 being of the
     "gamma-ray" type. Figure {\bf \ref{age_dist}} (left) demonstrates that "gamma-ray" pulsars in that catalogue
     cover a good range of distances in the Galaxy, and of ages.

    \subsection{Characteristics of the pulsar electron emission}
     
     The measurement accuracy on rotation frequency ($\Omega = 2\pi/P$ [Hz]) and frequency derivative give an
     estimation on the pulsar age. A pulsar can be modeled like a rotating neutron star with a dipolar magnetic
     field misaligned with its rotation axis, producing pulsed radiation from the magnetic poles.In this way,
     the electromagnetic energy comes from rotational energy following a braking law of $\dot{\Omega} \propto -\Omega^3$,
     implying an energy variation $\dot{E} = I\Omega\dot{\Omega}$, proportional to the moment of inertia $I$.
     The pulsar rotational velocity can be written as {\bf \cite{longair_2}}

     \begin{equation}
        \Omega(t) = \frac{\Omega_0}{(1+t/\tau_0)^{1/2}}
     \end{equation} 

     where $\Omega_0$ is the initial spin frequency and $\tau_0$ a decay time expressed as

     \begin{equation}
         \tau_0 = \frac{3c^3 I}{B^2R_s^6\Omega_0^2}
     \end{equation}

     where $I$ is the moment of inertia of the pulsar. $\tau_0 \sim 10^4$ years for nominal parameters
     {\bf \cite{aharo_tau0}}. $\dot{E}$ is compared to the pulsar age in figure {\bf \ref{age_dist}} (right),
     which establishes that the oldest objects have the lowest $\dot{E}$. At the same time, for each age 
     sample, "gamma-ray" ones have the higher $\dot{E}$. Mature pulsars could contribute to $e^{\pm}$ flux
     because produced particles are no longer trapped {\bf \cite{chi_1996}}. This framework allows us to
     derivate an estimated total released energy from a pulsar. Indeed, the upper limit to the rate of energy
     for electron-positron pairs is given by
    
     \begin{equation}
       \dot{E} = I \Omega \dot{\Omega} = \frac{1}{2}I \Omega^2_0 \frac{1}{\tau_0} \frac{1}{(1+\frac{t}{\tau_0})^2}
     \end{equation}
    
\begin{figure}[ht!]
       \begin{minipage}[c]{.45\linewidth}
         \centering
         \includegraphics*[height=7.5cm,width=7.5cm]{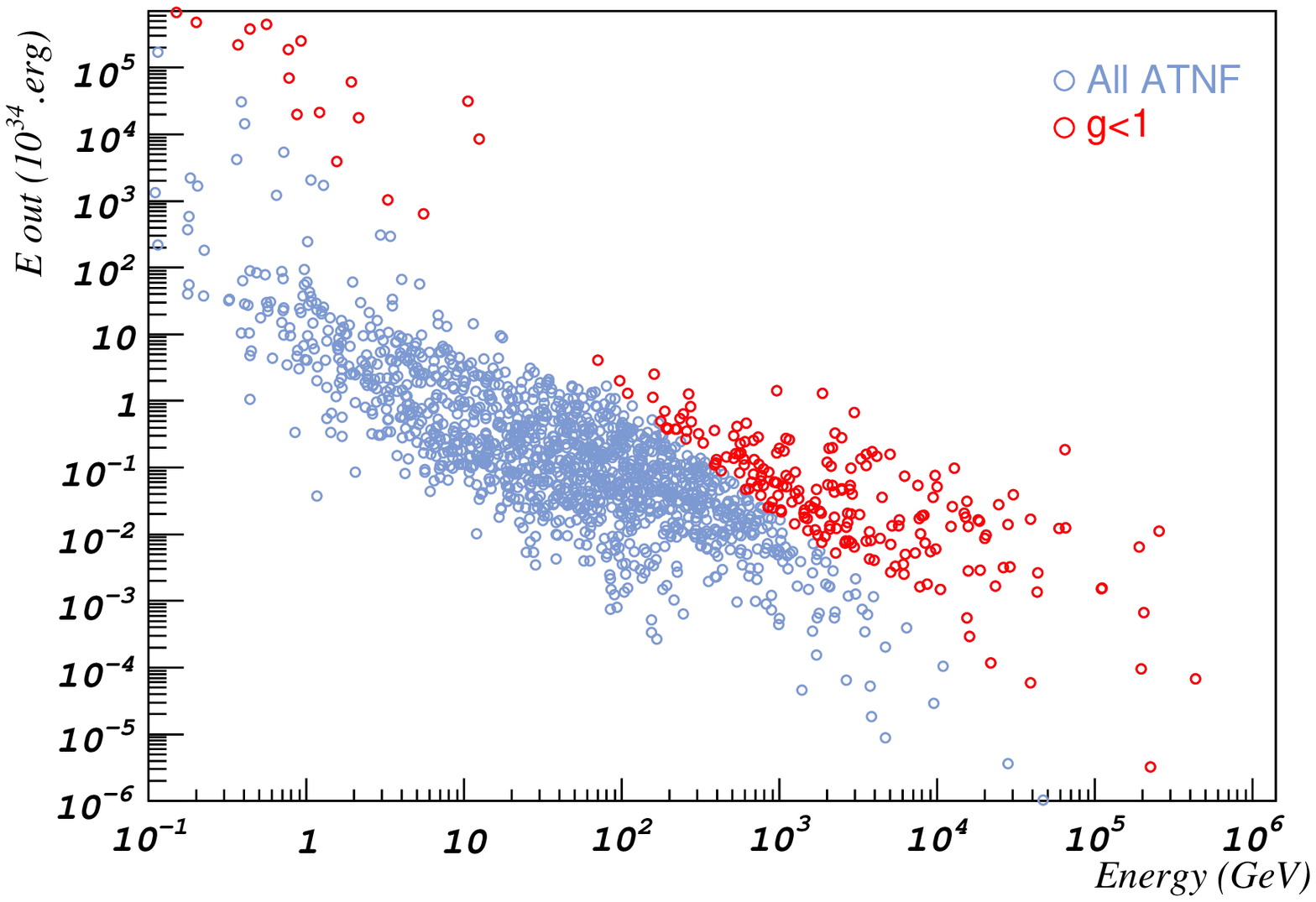}
       \end{minipage}\hspace{0.8cm}
       \begin{minipage}[c]{.45\linewidth}
         \centering
         \includegraphics*[height=7.5cm,width=7.5cm]{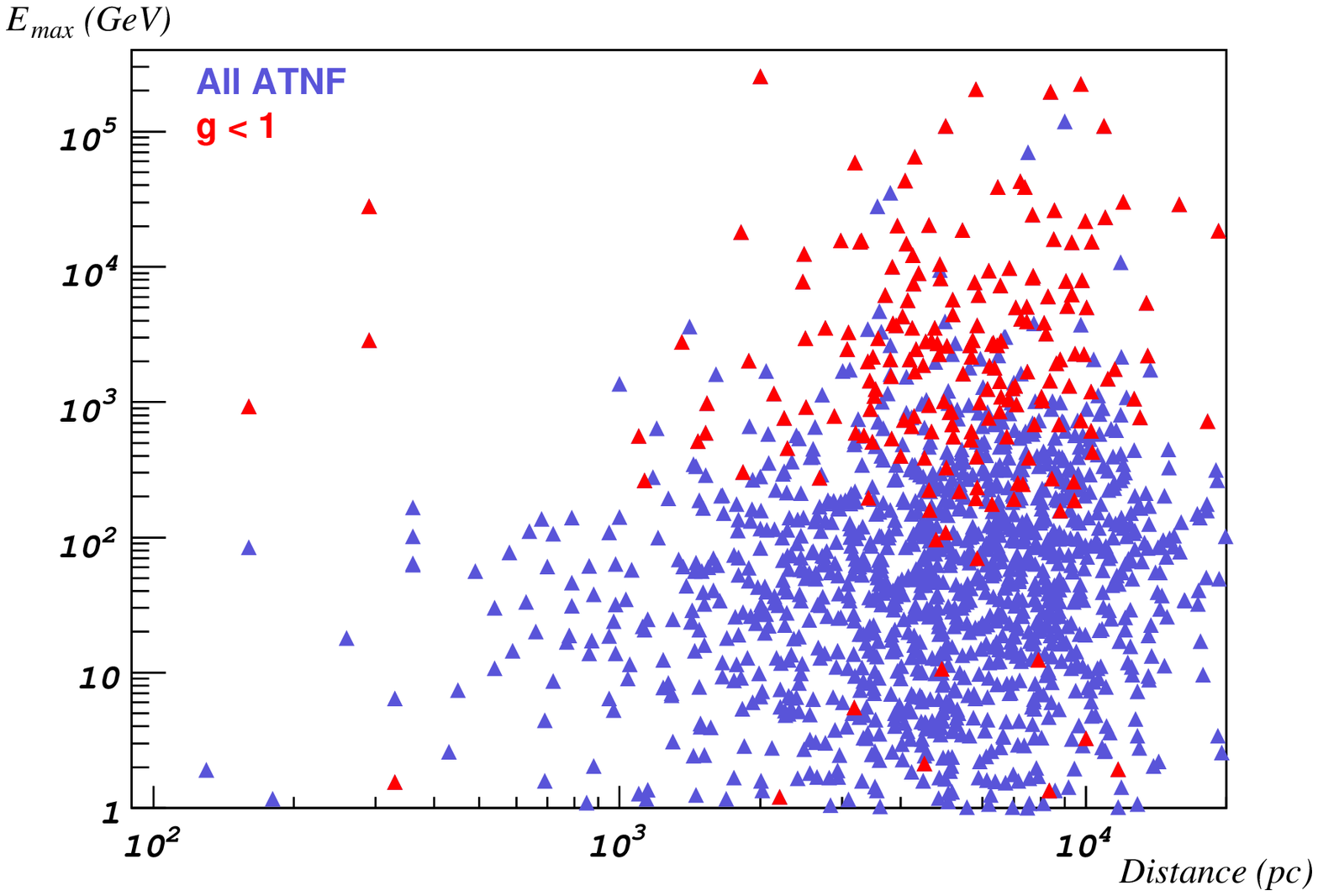}
       \end{minipage}
       \caption{\it Left:  Standard pulsar $E_{out}^{e^{\pm}}$ versus $E_{max}$ reached. 
	               Right: $E_{max}$ reached by pulsar versus pulsar distance.}
       \label{eout_age}
     \end{figure}

     From this equation, assuming an efficiency factor $f_{e^{\pm}}$ for $e^{\pm}$ pair production, the total
     energy that a mature pulsar (t $\gg$ $\tau_0$) has injected in magnetic dipole radiation is {\bf \cite{hooper_08}}
     {\bf \cite{profumo}}
    
     \begin{equation}
        E_{tot} \approx \frac{f_{e^{\pm}}}{2} I \Omega_0^2 \approx f_{e^{\pm}} \dot{E} \frac{T^2}{\tau_0}
     \end{equation}
    
     where $T$ is the typical age and $\dot{E}$ the spin-down power. The expected value for the efficiency
     factor is $f_{e^{\pm}} \sim$ few \% {\bf \cite{hooper_08}}. Other models, like Harding and Ramaty
     {\bf \cite{harding}}, Chi, Cheng and Young {\bf \cite{young}}, and Zhang and Cheng {\bf \cite{zhang_2001}},
     provide another $E_{out}$ expression. As a consequence of energy losses from the source to the earth, the
     maximum energy of electrons reaching the observer is expressed as $E_{max} = 3.10^3/t_5$ $GeV$, with $t_5$
     representing the electron age in $10^5$ years unit. This implies that after $10^5$ years, the maximum electron
     energy will be 3 $TeV$. $E_{max}$ indicates the maximum reach by the pulsar electron spectrum. As a direct
     consequence, the oldest the pulsar, the lowest is the $E_{max}$. Figure {\bf \ref{eout_age}} (left) illustrates,
     for the whole ATNF catalogue, that $E_{out}^{e^{\pm}}$ decreases with $E_{max}$, and that most of "gamma-ray"
     pulsars have a spectrum component above 10-100 GeV. Figure {\bf \ref{eout_age}} (right) indicates that only 
     three of the "gamma-ray" pulsars inside 1 kpc radius are able to contribute above 100 GeV.
    
     The spectral shape is limited at the energy range by $E_{max}$ but also because $e^{\pm}$ cannot be accelerated
     to arbitrarily high energies and a cut-off is expected around the TeV scale. Below that, we will assume a power
     law $Q(E_e) \propto E^{-\alpha}$. EGRET {\bf \cite{thom_egret}} {\bf \cite{fierro_egret}} observations of galactic
     pulsars give a power index $\alpha$ for gamma-ray spectra between 1.4 and 2.2. Assuming the power law measurement
     of cosmic electrons(positrons) of around 3.1 (3.3) {\bf \cite{barwick}}, a pulsar contribution with enough intensity
     is expected to appear in the electron spectrum. Table {\bf \ref{table_pulsar}} summarizes some properties for a
     sample of pulsars and supernova remnant Loop I {\bf \cite{profumo}} {\bf \cite{kobayashi}} which will be represented
     like a pulsar. All of the objects being close enough and with a large spin-down energy loss, giving total energy
     released $E_{tot}^{out}$. In order to contribute above 10 GeV, these pulsars must be close-by ($\sim$ 100-1000 $pc$)
     and their age must be less than {$3\cdot10^7$}  years. Geminga and Monogem are the two most probable and/or popular
     sources for the positron fraction excess. The two are experimentally interesting because of the intensity and the
     spectrum range.
     
     \begin{table}[ht!]
       \caption{ \it Data for selected nearby pulsars.}     
       \begin{center}
        \begin{tabular}{l|rcc|cc}
         \hline
         \hline
	 Name                 & Dist. &   Age          & g    & $E_{tot}^{out}$ & $E_{max}$\\
	                      &  (pc) &  (years)       &      & ($10^{50}$ GeV) & (GeV)\\
         \hline
         \hline
	 Geminga [J0633+1746] & 160.  &  3.42 $10^{5}$ & 0.70 &  74.            & 930.\\
	 Monogem [B0656+14]   & 290.  &  1.11 $10^{5}$ & 0.14 &   9.            & 2850.\\
	 Vela [B0833-45]      & 290.  &  1.13 $10^{4}$ & 0.70 &  17.            & 28040.\\
	 B0355+54             & 1100. &  5.64 $10^{5}$ & 0.61 &  281.           & 562.\\
         \hline
	 Loop I [SNR]         & 170.  &  2.   $10^{5}$ & -    &  43.            & 1584.\\
        \hline
        \hline
        \end{tabular} 
       \end{center}
       \label{table_pulsar}
     \end{table}

     From this short-list, one can understand the importance of the energy range for the detector. Since pulsars can
     contribute from GeV to TeV scale, it is crucial to see a cut-off and/or decrease to validate the pulsar
     contribution. For example, PAMELA seems to be too low in energy to detect such cut-off, while AMS-02 and Fermi-LAT
     with their extended range up to TeV may detect it.

      
  \section{Flux at Earth after propagation}

    \subsection{Solution of diffusion equation}
     
     After production of $e^{\pm}$, these particles mostly loose energy by synchrotron radiation (SR) and
     inverse Compton scattering (ICS). Their motion depends on the galactic magnetic field, which
     enables the direction reconstruction for charged particles. These processes are calculated by
     solving the transport equation in the standard diffusion approximation (neglecting convection),
     which for local sources can be expressed with a spherical symmetry, and reduced to the form
     {\bf \cite{atoyan_pro}}
     
     \begin{equation}\label{eq_propa}
       \frac{\partial}{\partial t} \frac{dn_e}{dE_e} = \frac{K(E_e)}{r^2} \frac{\partial}{\partial r} 
                       \left[r^2 \frac{\partial}{\partial r} \frac{dn_e}{dE_e}\right]
		     + \frac{\partial}{\partial E_e} \left[b(E_e)\frac{dn_e}{dE_e}\right] + Q(E_e)
     \end{equation}
     
     Here, $dn_e/dE_e$ is the number density of $e^{\pm}$ per unit energy, $K(E_e)$ is the diffusion parameter,
     $b(E_e)$ is the rate of energy loss, and $Q(E_e)$ the source term. We assume $K(E_e) \equiv 
     K_0(E_e/1 \textrm{GeV})^{\gamma}$, with $K_0$ and $\gamma$ specified in table {\bf \ref{table_diff}},
     where scenario MAX maximizes the positron flux and MIN minimizes it, and $b(E_e) = -bE^2_e$ with
     $b = 10^{-16} \textrm{GeV}^{-1}\textrm{s}^{-1}$. 

     \begin{table}[ht!]
       \caption{\label{table_diff} \it Different scenarios for diffusion parameters {\bf \cite{delahaye}}.}     
       \begin{center}
        \begin{tabular}{l|cc}
         \hline
          Scenario & $K_0$ & $\gamma$ \\
         \hline
          MAX & 1.8 $10^{27}$ & 0.55\\
          MED & 3.4 $10^{27}$ & 0.7\\
          MIN & 2.3 $10^{28}$ & 0.46\\
         \hline
        \end{tabular} 
       \end{center}
     \end{table}
             
     Assuming a power law injection, the solution of equation {\bf \ref{eq_propa}} is given by
     {\bf \cite{atoyan_pro}}

     \begin{equation}\label{dne_dee}
       \frac{dn_e}{dE_e} = \frac{Q(E_e)}{\pi^{3/2}r^3} (1-btE_e)^{\alpha-2}\left(\frac{r}{r_{diff}}\right)^3
                                       e^{(r/r_{diff})^2}
     \end{equation}     
     
     where $E < E_{max} \approx 1/(bt)$, and by $dn_e/dE_e = 0$ otherwise, with $r_{diff}$ being
     
     \begin{equation}
        r_{diff} \approx 2 \sqrt{K(E_e)t\frac{1-(1-E/E_{max})^{1-\gamma}}{(1-\gamma)E/E_{max}}}
     \end{equation} 
   
\begin{figure}[ht!]
      \begin{minipage}[c]{.45\linewidth}
        \centering
        \includegraphics*[height=7.5cm,width=7.5cm]{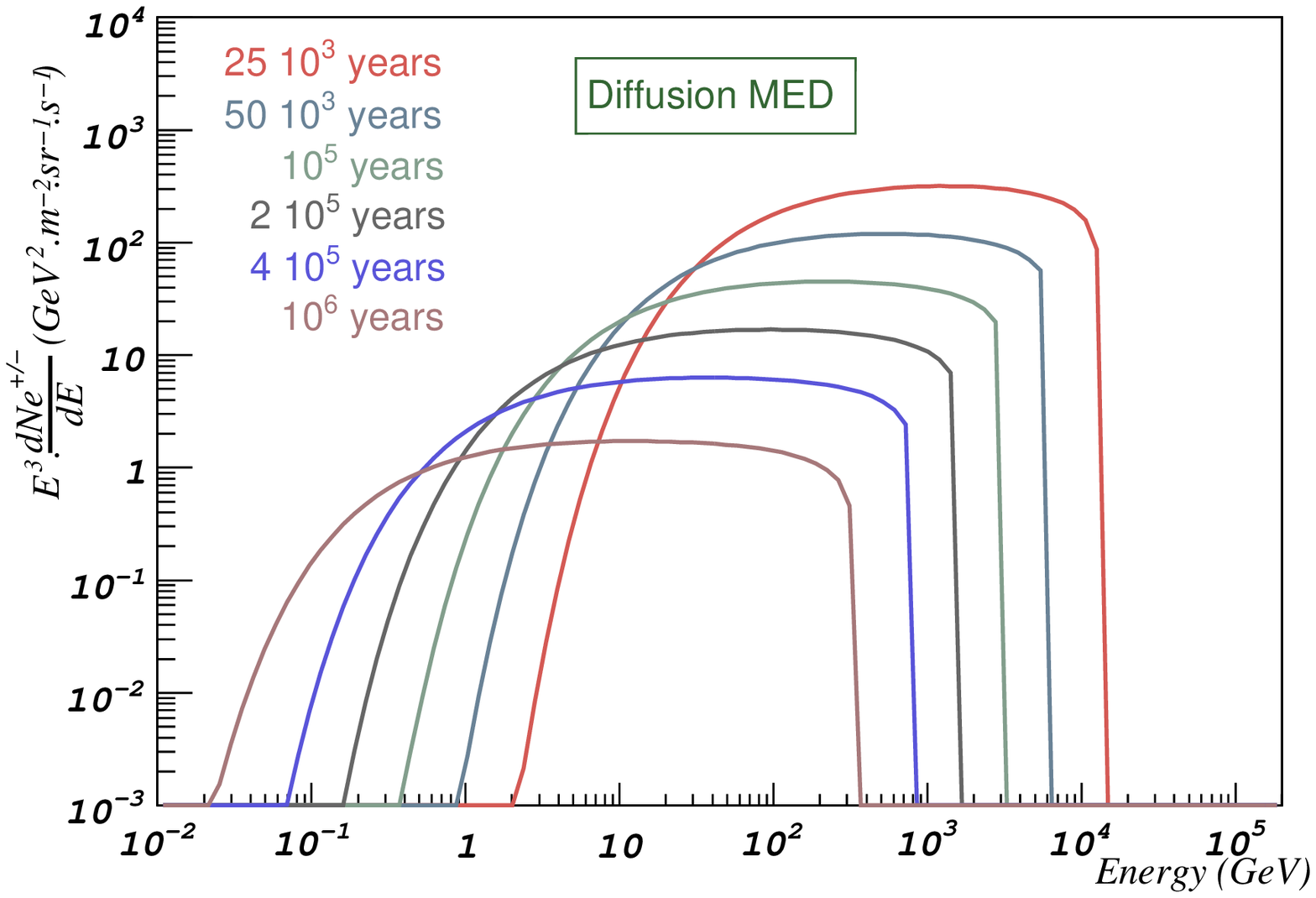}
      \end{minipage}\hspace{0.8cm}
      \begin{minipage}[c]{.45\linewidth}
        \centering
        \includegraphics*[height=7.5cm,width=7.5cm]{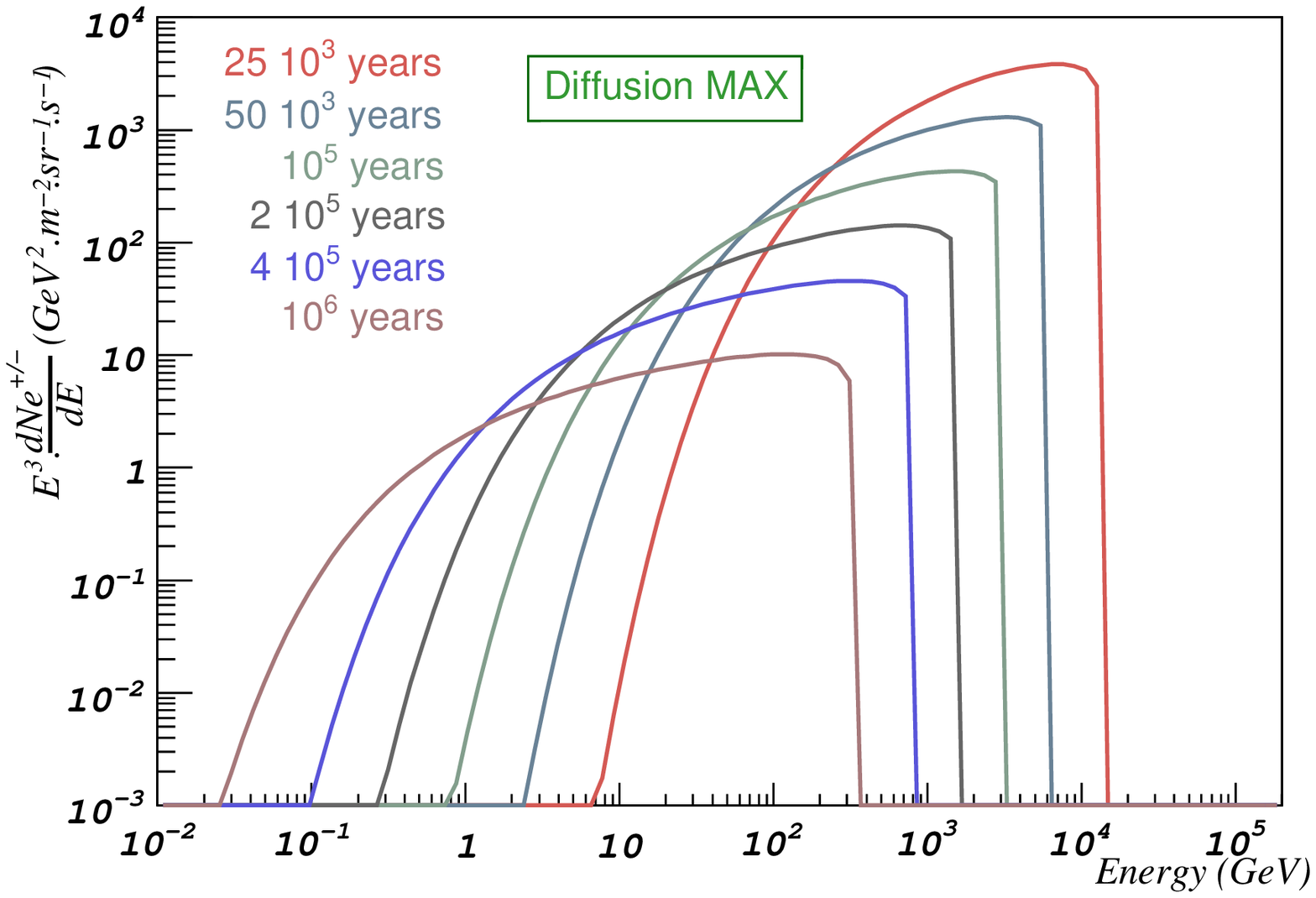}
      \end{minipage}
        \caption{\it Energy spectra of electrons at different injection time from a source at r = 160 pc,
	             $\alpha = 2$, and $E_{out} = 6\cdot10^{50}$ GeV for two diffusion scenarios:MED (left) and
		     MAX (right) (see  tab. {\bf \ref{table_diff}}).}
        \label{pulsar_age}
     \end{figure}

     The local $e^{\pm}$ flux from pulsars is fixed by pulsar distance $r$, the time of injection $t$
     which can be the pulsar age $t = T$ for mature pulsars, and the normalization of the injected flux.
     In this way, the local flux is just $J(E_e) = c (dn_e/dE_e) / (4\pi)$. Figure {\bf \ref{pulsar_age}}
     establishes energy spectra for different injection times $t$ and two different diffusion scenarios
     (see tab. {\bf \ref{table_diff}}). From these figures, one can see that contributions from old pulsars
     can be negligible compared to young ones; at the same time, for young sources, the diffusion time
     limits the contribution. Moreover, one understands that propagation parameters are key for spectrum
     shape and intensity. Indeed, in the MED scenario (see tab. {\bf \ref{table_diff}}), contribution from
     low energy is relatively more important than for the MAX scenario, while the shape of the MAX energy
     spectra is more easily distinguishable than in the MED case. From the experimental detection point of
     view and in particular for AMS-02, it is clear that the MAX scenario is preferred because of the flux
     shape. In any case, having a strong cut-off at $E_{max}$ could be a way to identify each pulsar
     contribution. The next sub-section will present the contribution in positron fraction of a sample of
     close-by pulsars.    

    \subsection{Positron fraction from nearby pulsars}\label{pos_frac}

     As first experimental study, only the closest pulsars, described in table {\bf \ref{table_pulsar_prop}}, 
     will be assumed as sufficient sources to explain the positron excess. Electrons flux will be determined
     for each one using equation {\bf \ref{dne_dee}} with $\alpha=2$. Positron fraction is adjusted with
     these sources for typical propagation parameters set: MAX, MED and MIN presented in table {\bf \ref{table_diff}}.
     Results are shown in table {\bf \ref{table_pulsar}} where each pair production efficiency $f_{e^{\pm}}$ pulsar
     are adjusted to reproduce PAMELA positron fraction for all propagation scenarios. As expected, the scenario
     which maximizes electrons are performed with the lowest $f_{e^{\pm}}$. In this case, some $\%$ for $f_{e^{\pm}}$ are
     sufficient to explain positron fraction which is in agreement with expectations {\bf \cite{hooper_08}}. The other
     extremum case, MIN scenario, needs almost 30$\%$ of the total energy released by the pulsar, which seems unlikely.
     Values for $f_{e^{\pm}}$ are chosen arbitrary keeping some proportionality between propagation case, except for
     B0355+54 pulsar which have a strong $f_{e^{\pm}}$ to create an irregularity in the total contribution.
      
     \begin{table}[ht!]
       \caption{\label{table_pulsar_prop} \it Data for selected nearby pulsars. Pair production efficiency $f_{e^{\pm}}$
       are calculated for different propagation scenario (see table {\bf \ref{table_diff}}).}     
       \begin{center}
        \begin{tabular}{l|c|rcl}
         \hline
         \hline
	 Name                 & $E^{out}_{tot}$ & \multicolumn{3}{c}{$f_{e^{\pm}}$} \\
	                      & ($10^{50}$ GeV) & (MAX) & (MED) &  (MIN)\\
         \hline
         \hline
	 Geminga [J0633+1746] &  74.  & 0.04 & 0.15 & 0.33 \\
	 Monogem [B0656+14]   &  9.   & 0.03 & 0.10 & 0.25 \\
	 Vela [B0833-45]      &  17.  & 0.03 & 0.10 & 0.25 \\
	 B0355+54             &  281. & 0.30 & 0.30 & 0.40  \\
         \hline
	 Loop I [SNR]         &  43.  & 0.04 & 0.10 & 0.33 \\
        \hline
        \hline
        \end{tabular} 
       \end{center}
     \end{table}

\begin{figure}[ht!]
      \begin{minipage}[c]{.45\linewidth}
        \centering
        \includegraphics*[height=7.5cm,width=7.5cm]{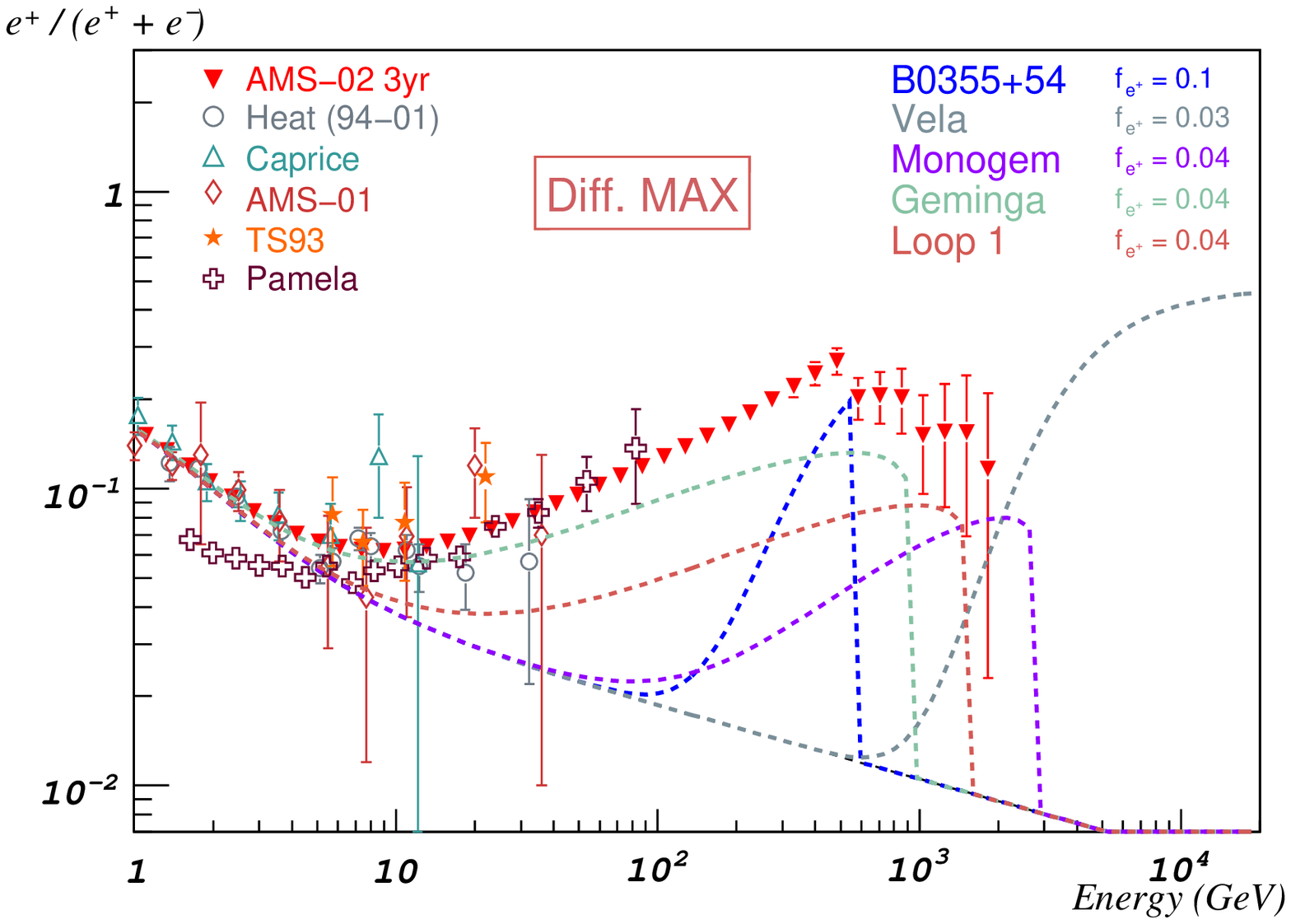}
      \end{minipage}\hspace{0.8cm}
      \begin{minipage}[c]{.45\linewidth}
        \centering
        \includegraphics*[height=7.5cm,width=7.5cm]{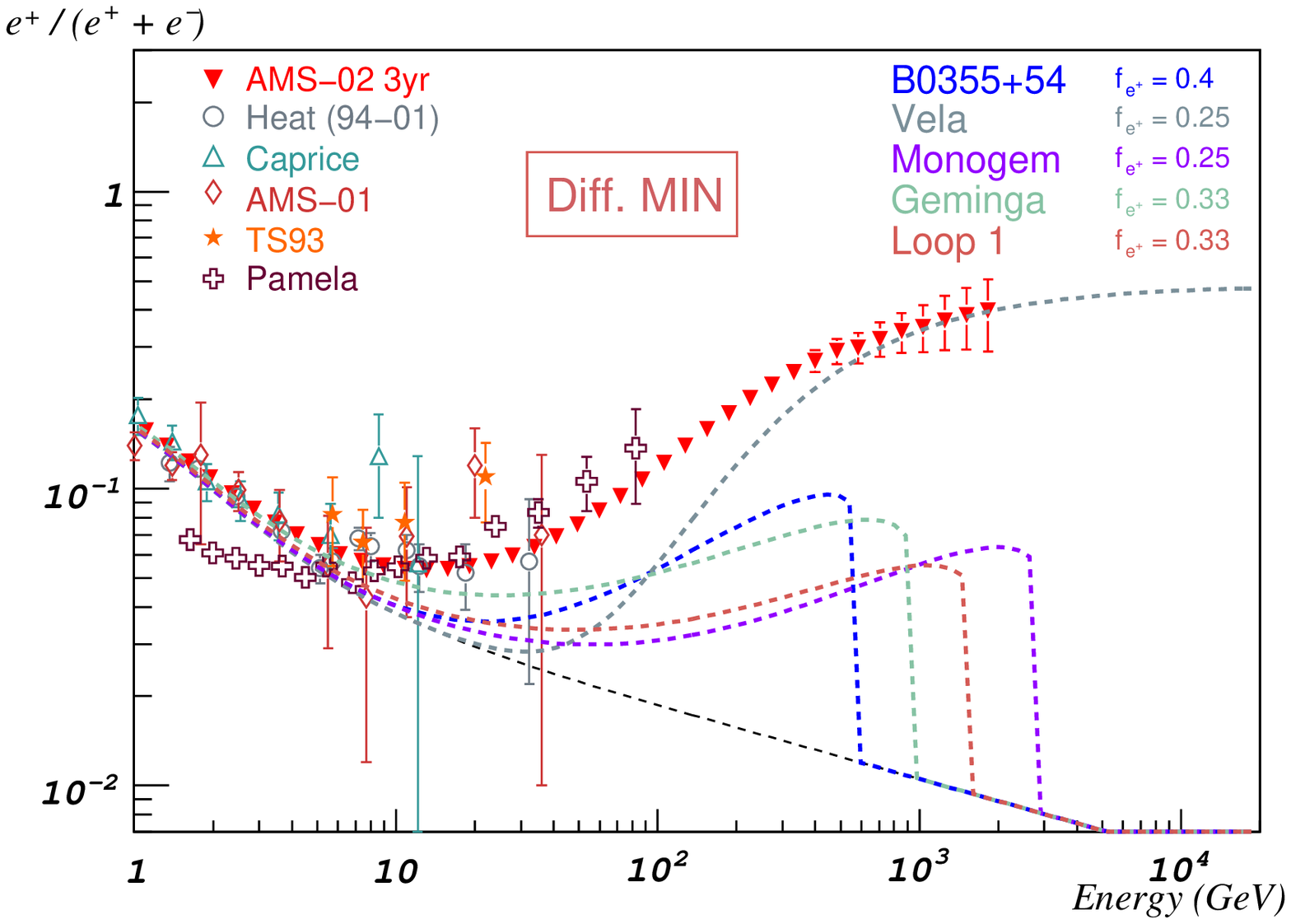}
      \end{minipage}
        \caption{\it Positron fraction reproduced by pulsar contributions ($\alpha=2$) for two
	             propagation scenarios with AMS-02 capacity: MAX (left) and
		     MIN (right) (see table {\bf \ref{table_pulsar_prop}}).}
        \label{combi_puls}
     \end{figure}

     Figures {\bf \ref{combi_puls}}  illustrate positron fraction prediction for AMS-02 for the extreme cases MAX scenario
     (left) and MIN scenario (right) using electrons and positrons background adjusted by Baltz et al. {\bf \cite{baltz_99}}.
     For these predictions, AMS-02 acceptance is taken to be $\mathcal{A}_{e^{\pm}}=0.045 m^2 \cdot sr$.
     Further informations on the AMS-02 electron/proton separation and its acceptance can be found in the literature
     {\bf \cite{john_these}} {\bf \cite{pos_ams}}. Two remarks can be made about AMS-02 electron acceptance. Firstly,
     below 10 GeV, the acceptance is lower than this mean value but the rate of cosmic rays is strong enough, so electron
     flux will still be high and will compensate the acceptance. Secondly, at energy above 500 GeV, the electron/proton
     separation will be more challenging for the detector, while at the same time the proton contamination is decreasing
     with energy, which could help to keep a good separation. For these reasons, the electron acceptance will be assume
     energy-independent. For the MAX scenario {\bf \ref{combi_puls}} (left), a structure from pulsars contribution can be
     detected by AMS-02 and a clear decreasing appears above 500 GeV.
      
     The MIN scenario, figure {\bf \ref{combi_puls}} (right), presents a total contribution without irregularity
     increasing above TeV scale. This scenario give a smooth contribution and it would be difficult to distinguish
     single pulsar contribution, even in the case of multiple $f_{e^{\pm}}$ values. The same problem comes with the
     MED scenario which gives also a continuous flux from pulsars. To conclude, propagation scenario MAX case gives
     the best experimental conditions with pair production parameters $f_{e^{\pm}}$ as expected by models up to a few
     percent. In this scenario, pulsars with a large $f_{e^{\pm}}$ (B0355+54 in our case) must appear like a peak in
     the spectrum or giving a clear cut-off detectable by AMS-02, like seen in figure {\bf \ref{combi_puls}} (left).
     On the other hand, the MIN scenario may smooth all contributions making almost impossible the pulsar distinction,
     and it implies important $f_{e^{\pm}}$, around 30 $\%$, to reproduce PAMELA data (figure {\bf \ref{combi_puls}}
     (right)), indicating that pulsars couldn't be the unique contribution to this excess. For this study, electrons
     background was assumed for a mean value, Delahaye et al. {\bf \cite{delahaye_09}} presented uncertainties for
     leptons background where background can be higher and, in this case pulsars flux may be lower.
 
    \subsection{Positrons flux from pulsars continuum}\label{fond_elec}
         
\begin{figure}[ht!]
      \begin{minipage}[c]{.45\linewidth}
        \centering
        \includegraphics*[height=7.5cm,width=7.5cm]{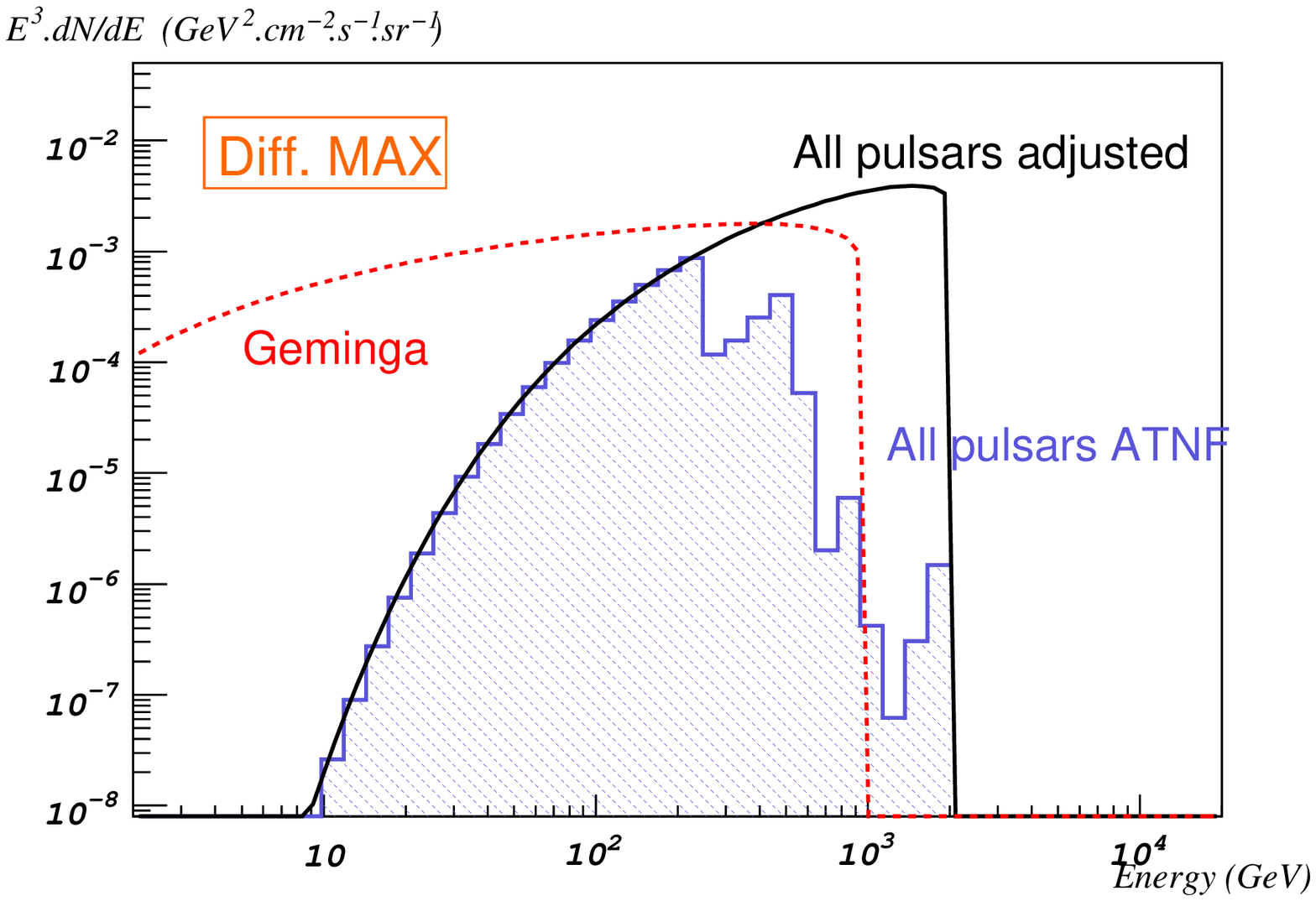}
      \end{minipage}\hspace{0.8cm}
      \begin{minipage}[c]{.45\linewidth}
        \centering
        \includegraphics*[height=7.5cm,width=7.5cm]{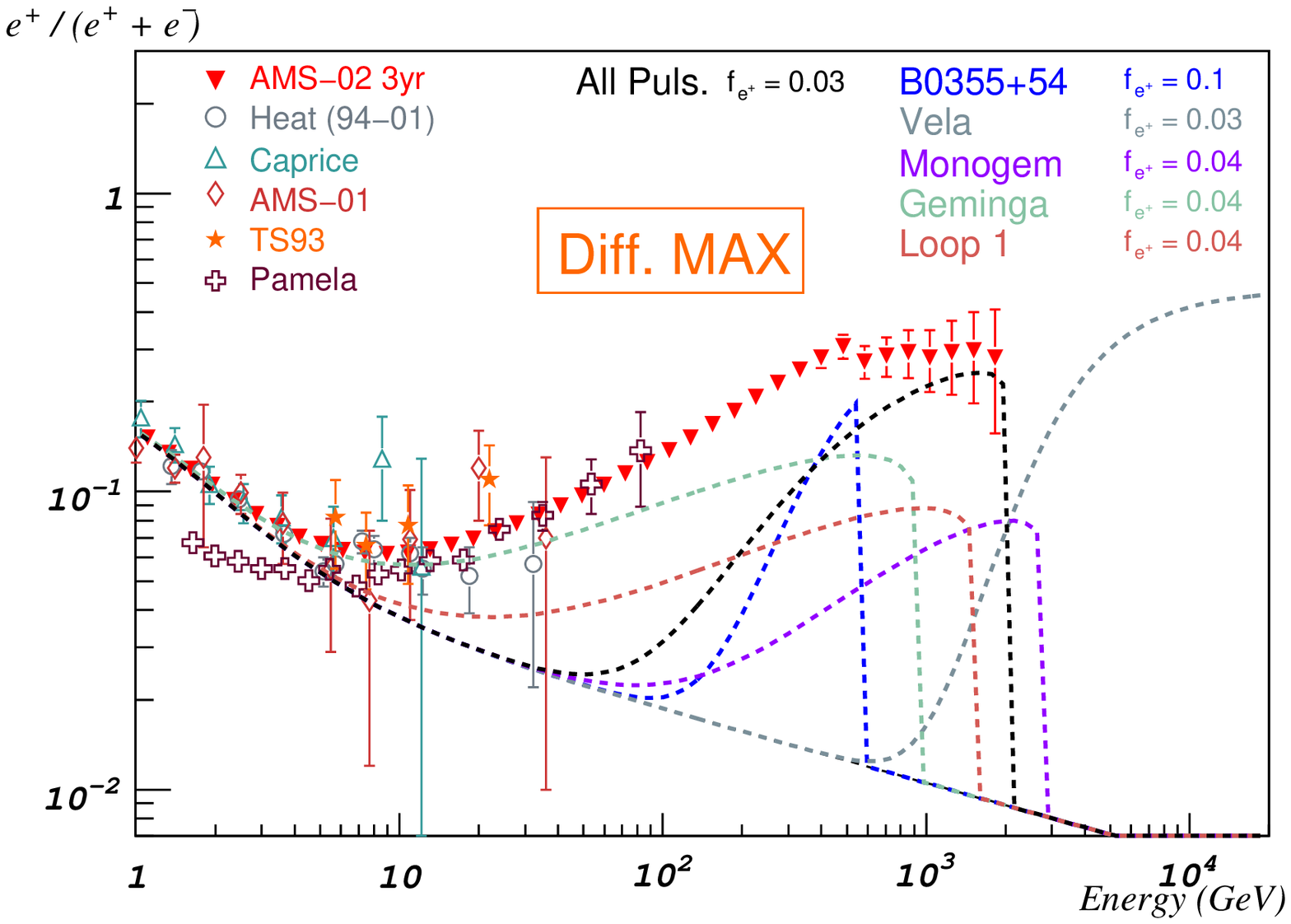}
      \end{minipage}
        \caption{\it Left: Electrons flux from all ATNF gamma-ray pulsars (r $>$ 1 kpc) compared to Geminga with $f_e^{\pm} = 0.03$.
	             Right: Positron fraction reproduced by nearby pulsars and pulsars continuum with AMS-02 capacity.}
        \label{pulsar_fond}
     \end{figure}

     In section {\bf \ref{pos_frac}}, only nearby pulsars had been considered sufadficient to reproduce positron fraction. 
     ATNF catalogue gives 184 more "gamma-ray" type pulsars above 1 kpc, and an age compatible with a contribution above
     10 GeV. To determine a contribution for all these pulsars, the same $f_{e^{\pm}}$ is assumed. Pulsars continuum is
     calculated with $f_{e^{\pm}} = 0.03$ in figure {\bf \ref{pulsar_fond}} (left) for the MAX scenario. The total
     contribution have similar electrons production to Geminga pulsar, in the same condition. A simple way to represent
     the whole faraway pulsars could be to use a pulsar-like adjustment assuming a cut-off above hundreds GeV. In this
     section, a cut-off around 2 TeV will be assumed like extremum case where pulsars continuum can contribute at high
     energies. According the distance, pulsars continuum may contribute mostly above 50 GeV. Others studies {\bf \cite{hooper_08}} {\bf \cite{zhang_2001}}
     based on random distribution in the galaxy for a giving pulsar birthrate give contribution at lower energies because
     they are considering all pulsars even old ones, which are out this study. Indeed, we consider pulsars able to
     participate at the positrons excess at high energy.
        
     Pulsars adjustment from figure {\bf \ref{pulsar_fond}} (left) is used in same conditions with nearby pulsars to
     reproduce PAMELA positron fraction in figure {\bf \ref{pulsar_fond}} (right). Positrons from faraway pulsars should
     diminish structure effects and cut-off in the positron fraction. Above 500 GeV, continuum should hide individual
     pulsar giving plateau-like with a low decreasing rate. In the best case, choosing a lower cut-off for the continuum,
     in between 700 GeV to 1 TeV, will give a similar shape with a decreasing a bit bigger without cut-off. Assuming same
     conditions for all pulsars, continuum should participate to positron fraction at high energy diminishing detector
     capacity to determine primary sources. In the continuum case, AMS-02 should observe at least a plateau above 500 GeV
     with a slow decreasing.
     

  \section{Anisotropy: strong evidence for pulsar contribution?}\label{anis_elec_pos}
  
     The propagation of positrons-electrons does not allow to pinpoint each possible $e^{\pm}$ source, reducing
     the chance of constraining models, for example, which sources are responsible for the excess from dark matter.
     One possibility proposed by Mao \& Shen {\bf \cite{mao_shen}} is that the closest pulsars can induce an
     anisotropy in the pulsars direction. Although propagation affects electron trajectory, close-by pulsars can induce
     a small dipole anisotropy, which should be present at sufficiently high energy {\bf \cite{mao_shen}}. Anisotropy
     will be discussed for electrons-positrons and positrons case.
        
    \subsection{Anisotropy in the $e^+e^-$ flux}
     
     In a very general way, the anisotropy of the $e^{\pm}$ flux associated with diffusive propagation can be 
     calculated as {\bf \cite{ginzburg}}
     
     \begin{equation}\label{delta}
       \delta = \frac{I_{max}-I_{min}}{I_{max}+I_{min}} = \frac{3K|\nabla(dN_e/dE_e)|}{c(dN_e/dE_e)}
     \end{equation}
      
     where $\nabla(dN_e/dE_e)$ is the gradient of $e^{\pm}$ density. In the case of energy-independent diffusion
     anisotropy, Mao \& Shen {\bf \cite{mao_shen}} proposed an estimation of the maximum expected anisotropy as
      
     \begin{equation}
         \delta_{max} = \frac{3}{2c}\frac{r}{t}
     \end{equation}
       
     where $r$ and $t$ are respectively the distance and age of the pulsar. This simple relation forces us
     to choose not only the closest pulsars, but also young ones. Table {\bf \ref{anis_max}} shows some expected
     energy-independent pulsar anisotropies. Vela being a young pulsar gives a strong anisotropy with
     respect to Geminga or Monogem.
       
     \begin{table}[ht!]
       \caption{\label{anis_max} \it Maximum anisotropy for selected nearby pulsars.}     
       \begin{center}
        \begin{tabular}{l|cc|c} 
         \hline
         \hline
	 Name                 & Dist. (pc) &   Age (years)  & $\delta_{max}$ (\%)\\
         \hline
         \hline
	 Geminga [J0633+1746] & 160.       &  3.42 $10^{5}$ &  0.23\\
	 Monogem [B0656+14]   & 290.       &  1.11 $10^{5}$ &  1.28\\
	 Vela [B0833-45]      & 290.       &  1.13 $10^{4}$ &  12.5\\
	 B0355+54             & 1100.      &  5.64 $10^{5}$ &  0.95\\
	 Loop I [SNR]         & 170.       &  5.64 $10^{5}$ &  0.15\\
        \hline
        \hline
        \end{tabular} 
       \end{center}
     \end{table}

     For energy-dependent diffusion, inserting Eq. {\bf \ref{dne_dee}} in {\bf \ref{delta}}, anisotropy can be
     expressed as
      
     \begin{equation}\label{anis_dep}
        \delta = \frac{3}{2c} \frac{r}{t} \frac{(1-\gamma)E/E_{max}}{1-(1-E/E_{max})^{1-\gamma}} \frac{N_e^{Puls}}{N_e^{tot}}
     \end{equation}
      
     where $N_e^{Puls}$ and $N_e^{tot}$ are $e^{\pm}$ contribution from pulsars and from the electron
     background, taken from {\bf \cite{barwick}}. From the detection point of view, to observe an anisotropy
     at the 2$\sigma$ level, the observations must satisfy the condition  $\delta \geq 2\sqrt{2}/\sqrt{N_{evts}}$
     where $N_{evts}$ is the number of events collected above an energy threshold. 
     
     \begin{figure}[ht!]
      \begin{minipage}[c]{.45\linewidth}
        \centering
        \includegraphics*[height=7.5cm,width=7.5cm]{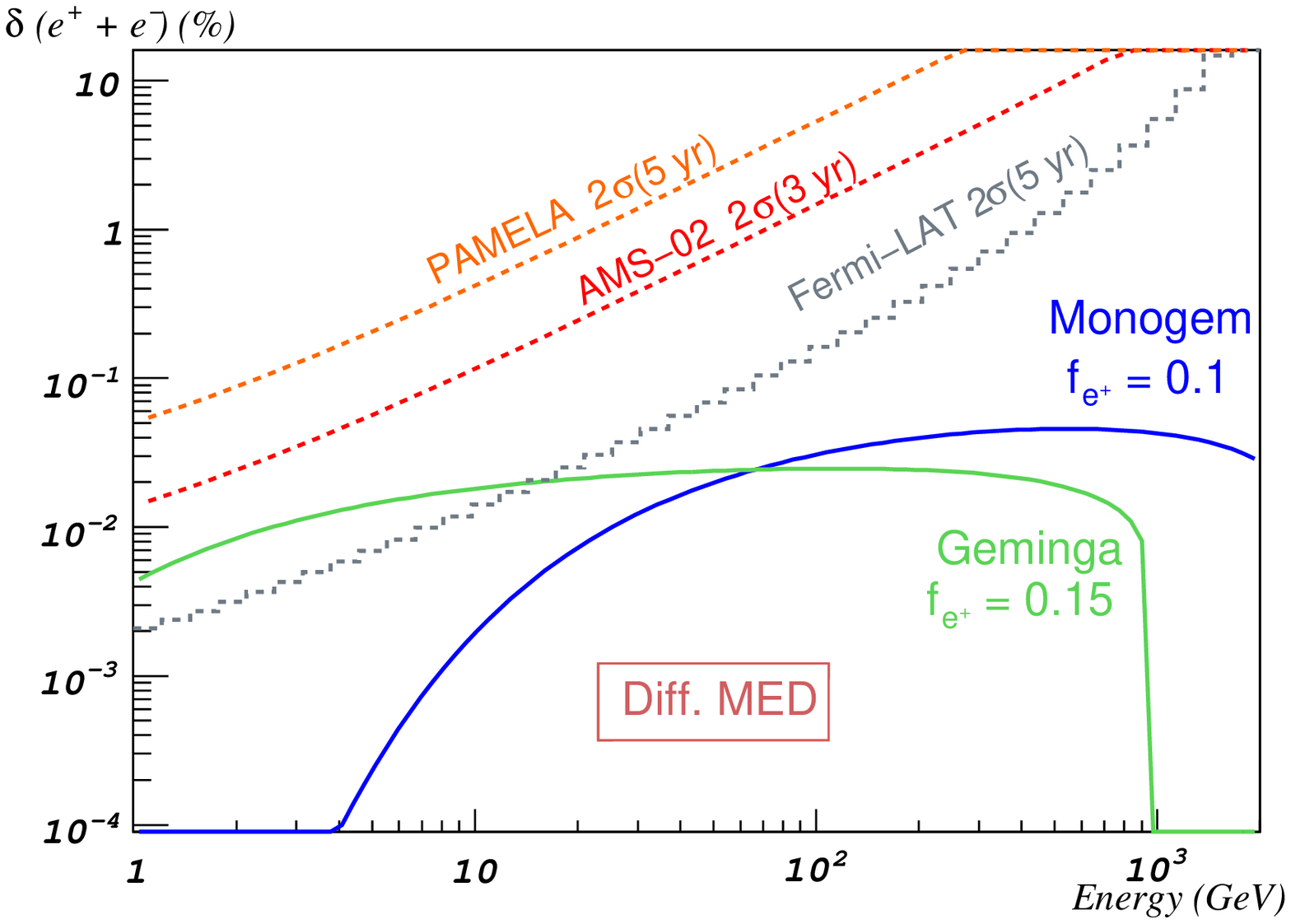}
      \end{minipage}\hspace{0.8cm}
      \begin{minipage}[c]{.45\linewidth}
        \centering
        \includegraphics*[height=7.5cm,width=7.5cm]{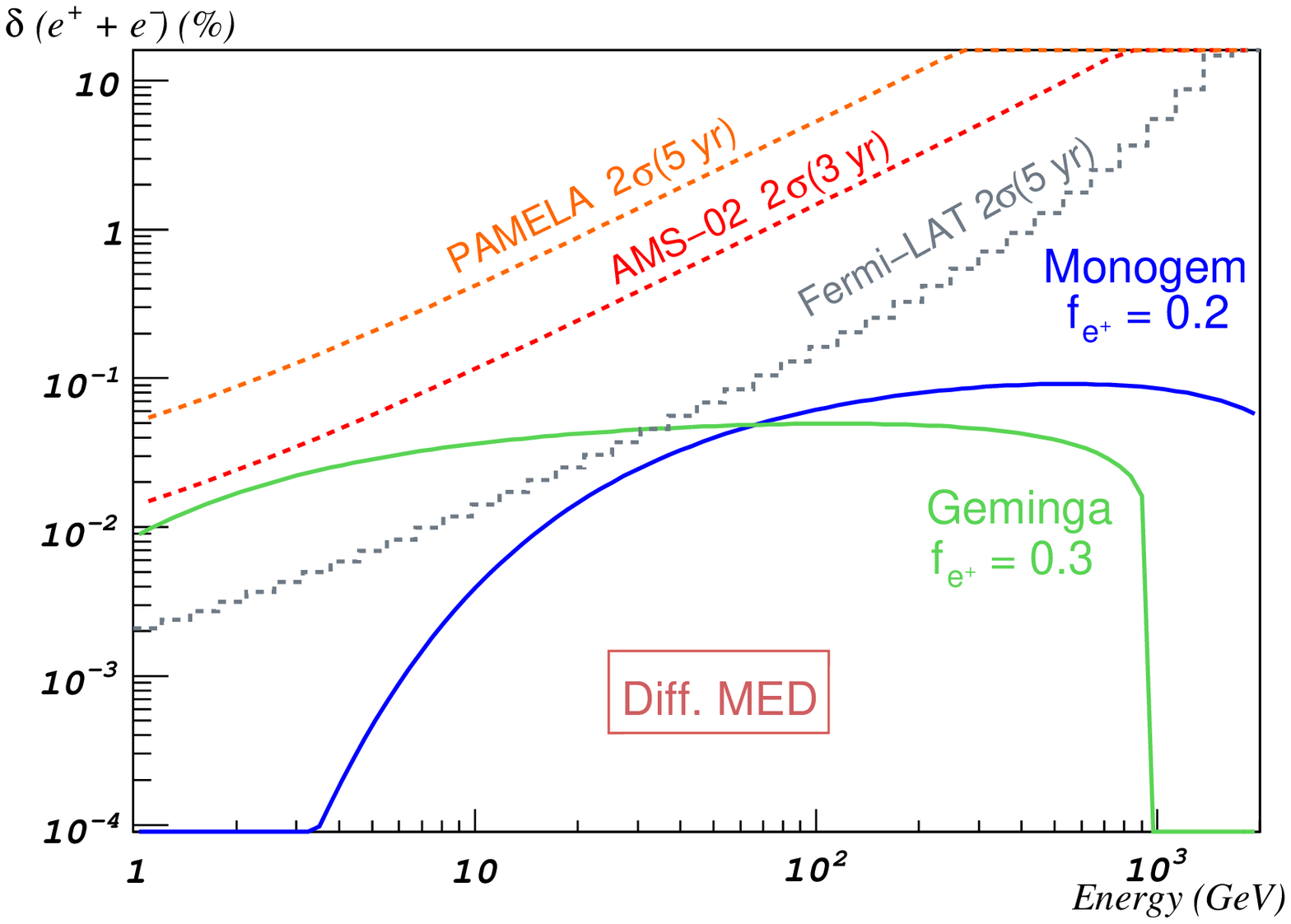}
      \end{minipage}
        \caption{\it Anisotropy in $e^{\pm}$ flux for Geminga and Monogem pulsars with sensitivity 
	             expected for AMS-02, PAMELA and Fermi-LAT experiments for MED propagation scenario 
		     (see table {\bf \ref{table_pulsar_prop}}). Geminga and Monogem are set to reproduce
		     alone positron excess (right), or with others pulsars (left) (see section \ref{pos_frac}).}
        \label{anis_tout}
     \end{figure}  

     Figures {\bf \ref{anis_tout}} show anisotropies expected for the two closest pulsars: Geminga and Monogem,
     where each pulsars flux is determined like in section {\bf \ref{pos_frac}}. In the figure 
     {\bf \ref{anis_tout}} (left), Geminga and Monogem are set like in the table {\bf \ref{table_pulsar_prop}}
     to adjust PAMELA positron fraction with others close pulsars, and figure {\bf \ref{anis_tout}} (left)
     presents a new configuration where Geminga and Monogem only contribute to positron excess with higher
     individual $e^{\pm}$ production. These anisotropies are represented for MED propagation scenario, and
     compared to the sensitivity expected of the experiments PAMELA, Fermi-LAT and AMS-02 with standard
     background. The main difference between experiments sensitivity is explained mostly by their electron
     acceptance: Fermi-LAT have $\mathcal{A}^{e^{\pm}}_{Fer} \sim 1.5-2$ $m^2 \cdot sr$ {\bf \cite{dat_fermi}},
     for AMS-02 it is $\mathcal{A}^{e^{\pm}}_{AMS} \sim 0.045$ $m^2 \cdot sr$ {\bf \cite{john_these}}, and for
     PAMELA, $\mathcal{A}^{e^{\pm}}_{PAM} \sim 0.002$ $m^2 \cdot sr$ {\bf \cite{exp_pamela}}. After five years,
     PAMELA will reach $\delta > 0.5\%$ at 2$\sigma$ above 10 GeV remaining too high to detect such anisotropy.
     AMS-02 and even more Fermi-LAT may be able to distinguish an anisotropy. For AMS-02 within three years,
     sensitivity should be the same order than Geminga anisotropy for the case where Geminga and Monogem have
     sufficient $e^{\pm}$ production to reproduce PAMELA data as it appears in figure {\bf \ref{anis_tout}} (left).
     To prevent the misunderstanding of anisotropy results, it would be better to look for anisotropy above 10
     GeV where starts the positron excess, for this reason looking for $e^{\pm}$ anisotropy, Fermi-LAT should be
     the best experiment with $\delta > 0.015\%$ at 2$\sigma$ above 10 GeV. Indeed in figure {\bf \ref{anis_tout}}
     (left), Fermi-LAT with five years statistic can establish an anisotropy for Geminga from low energy to 10 GeV
     in the multiple pulsars hypothesis. For the strong pulsars case, presented by {\bf \ref{anis_tout}} (right),
     Fermi-LAT should extend the accessible energy range for Geminga until 30 GeV, and may almost reach Monogem
     anisotropy. This study shows how it will be difficult for experiments to deal with anisotropy having sensitivity
     and anisotropy values very close. Besides the influence on the $f_{e^{\pm}}$, propagation will play a role
     in the spectrum pulsar shape and therefore should modify in the pulsar anisotropy. For the MAX scenario,
     contributions at low energy will diminish where most of the statistic are expected, and for the MIN one
     pulsars will give flat distribution and some anisotropy for Fermi-LAT below 10 GeV. For the $e^{\pm}$
     anisotropy, Fermi-LAT must be the best experiment to provide some anisotropy informations.

    \subsection{Anisotropy in $e^+$ flux: powerful detection}

     Electrons-positrons anisotropy is interesting because of high statistic reachable by Fermi-LAT experiment.
     An another possibility is to consider only positrons flux. Indeed, studying positron is a direct way to
     connect to the primary  $e^{\pm}$ source. Therefore, anisotropy in $e^+$ flux is the best way to characterize
     a close-by source and to constrain it, this study was well presented by Busching et al. {\bf \cite{busching}}.
     Fermi-LAT can not perform this measurement, because is not able to distinguish positrons from electrons. PAMELA
     and AMS-02 have both magnets to determine sign particle.  The average field inside the PAMELA magnet is 0.4 $T$
     {\bf \cite{exp_pamela}}, AMS-02 with its supraconductor magnet having a bending power of 0.86 $T \cdot m^2$
     {\bf \cite{ams_nim}}, will differentiate between electrons and positrons above 300 GeV.
           
     \begin{figure}[ht!]
      \begin{minipage}[c]{.45\linewidth}
        \centering
        \includegraphics*[height=7.5cm,width=7.5cm]{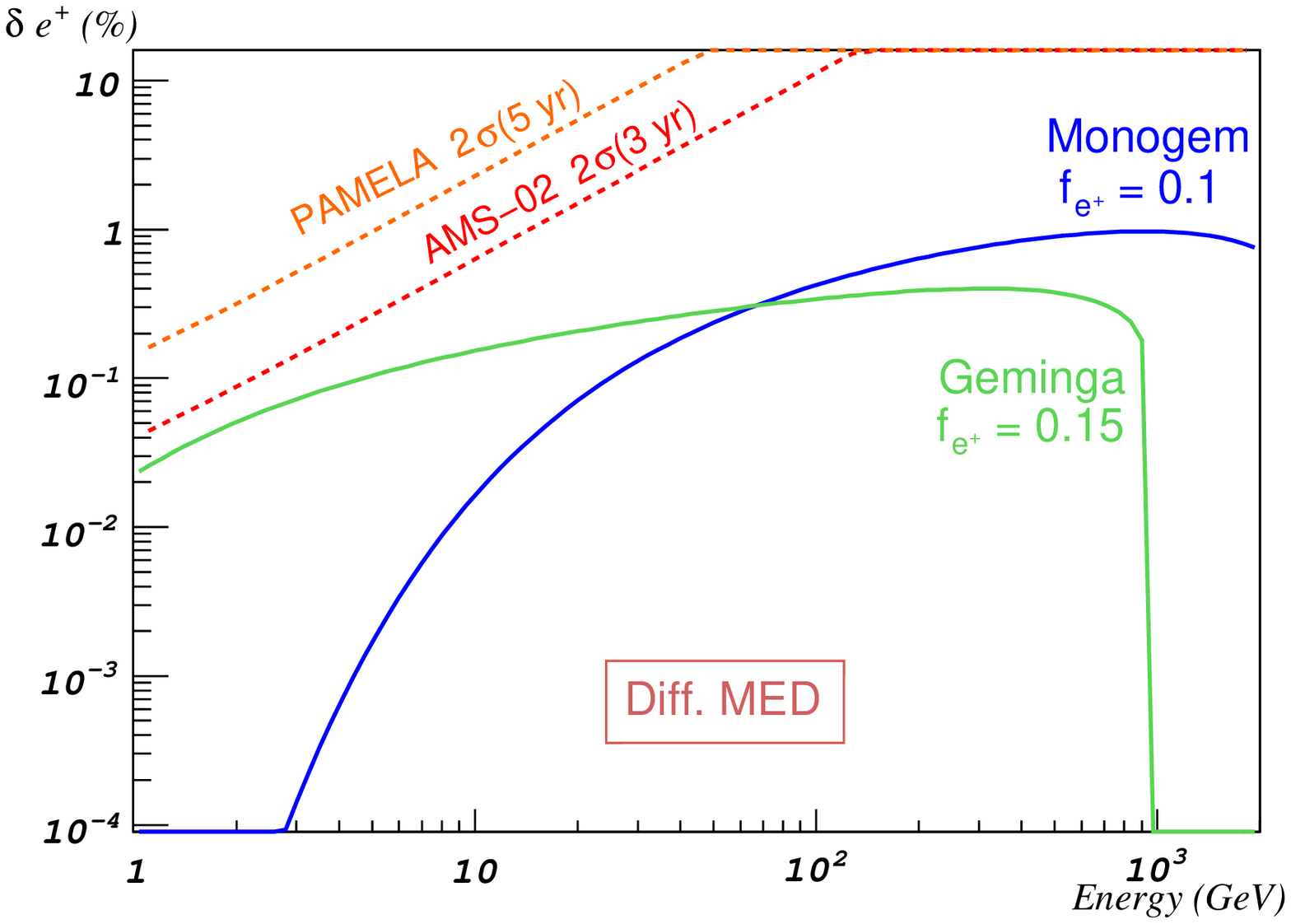}
      \end{minipage}\hspace{0.8cm}
      \begin{minipage}[c]{.45\linewidth}
        \centering
        \includegraphics*[height=7.5cm,width=7.5cm]{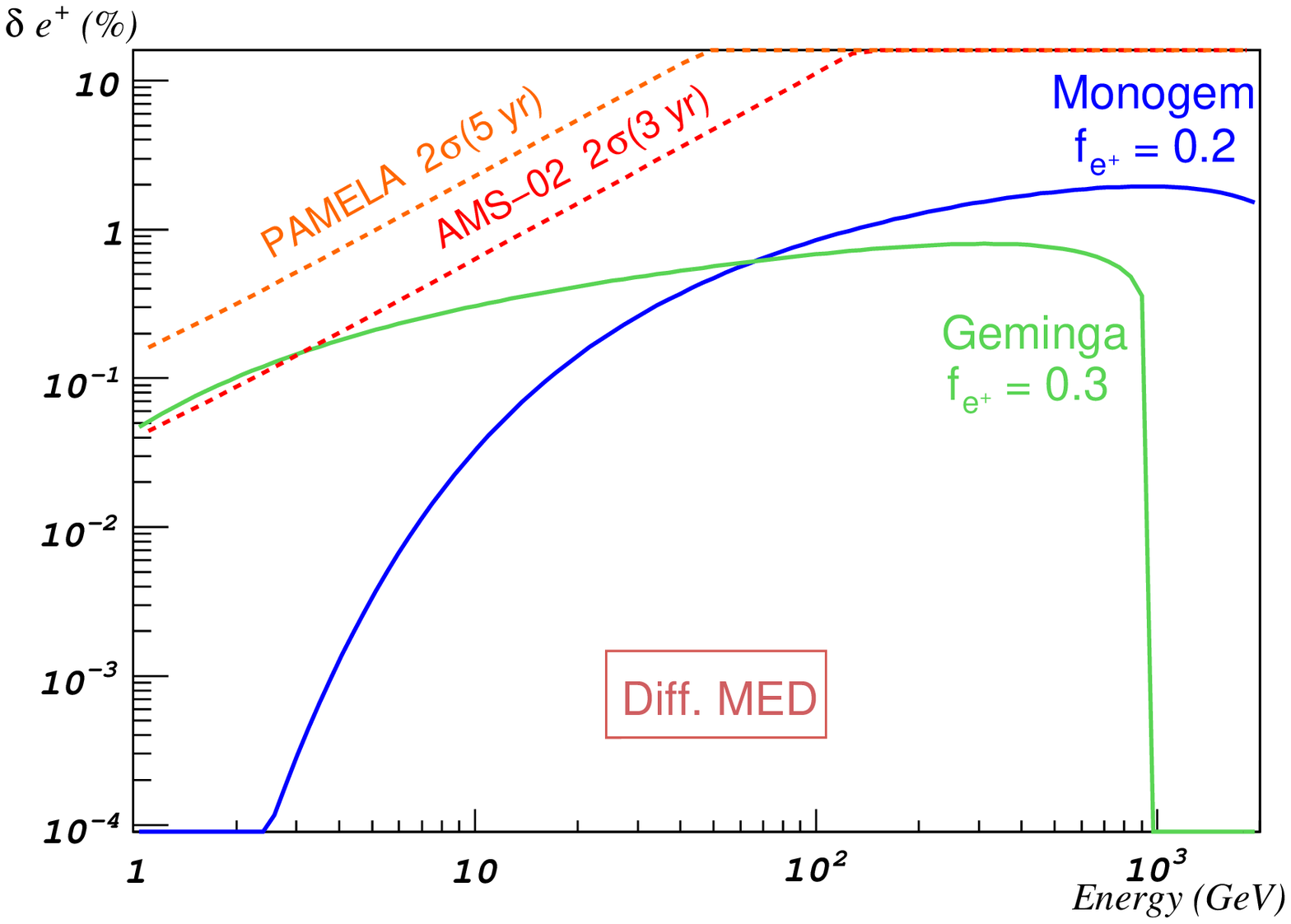}
      \end{minipage}
        \caption{\it Anisotropy in $e^+$ flux for Geminga and Monogem pulsars with sensitivity 
	             expected for AMS-02 and PAMELA experiments for MED propagation scenario 
		     (see table {\bf \ref{table_pulsar_prop}}).Geminga and Monogem are set to reproduce
		     alone positron excess (right), or with others pulsars (left) (see section \ref{pos_frac}).}
        \label{anis_pos}
     \end{figure}       
     
     The study performed with $e^{\pm}$, in section {\bf \ref{anis_elec_pos}}, is now applied to $e^+$. Figures 
     {\bf \ref{anis_pos}} present $e^+$ anisotropy for MED propagation with sensitivity at 2$\sigma$ level for
     PAMELA after 5 years and AMS-02 after 3 years. PAMELA will get $\delta > 0.1\%$ above 1 GeV after 5 years
     where $\delta$ must be around $0.01\%$ to see $e^+$ anisotropy. Figure {\bf \ref{anis_pos}} (left) confirms
     AMS-02 needs close pulsars with most of the energy transferred to $e^{\pm}$ to have sufficient discovery
     capacity like shown in figure {\bf \ref{anis_pos}} (right). AMS-02 is in best condition than in the previous
     section ({\bf \ref{anis_elec_pos}}), AMS-02 will be able to reach pulsar anisotropy below 10 GeV if few pulsars
     like Monogem and Geminga produce most of the $e^{\pm}$ excess. The difficulty for AMS-02 is to perform an
     anisotropy analysis below 10 GeV.
           
    \subsection{Discussion on anisotropy detection}
     
\begin{figure}[ht!]
      \begin{minipage}[c]{.45\linewidth}
        \centering
        \includegraphics*[height=7.5cm,width=7.5cm]{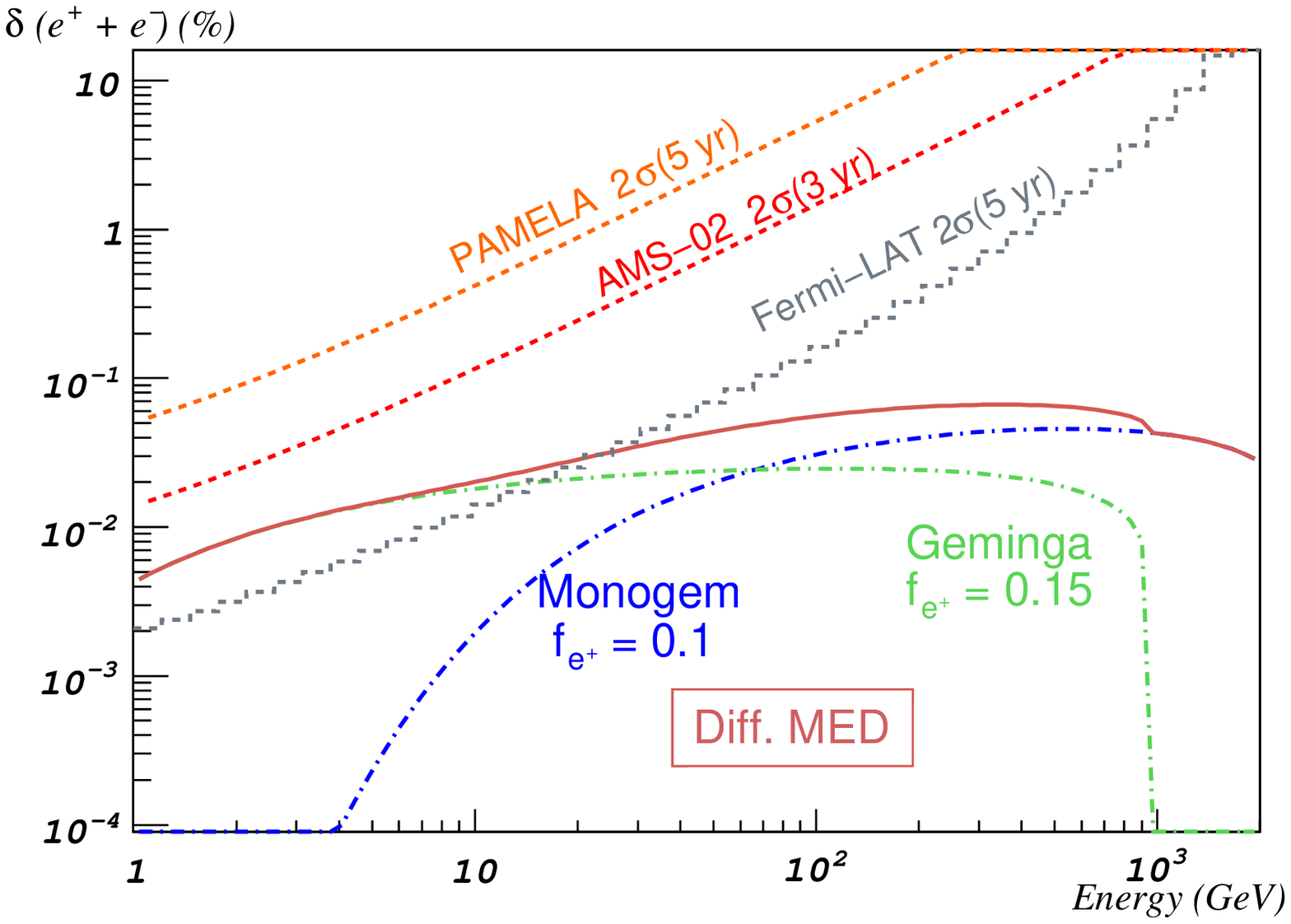}
      \end{minipage}\hspace{0.8cm}
      \begin{minipage}[c]{.45\linewidth}
        \centering
        \includegraphics*[height=7.5cm,width=7.5cm]{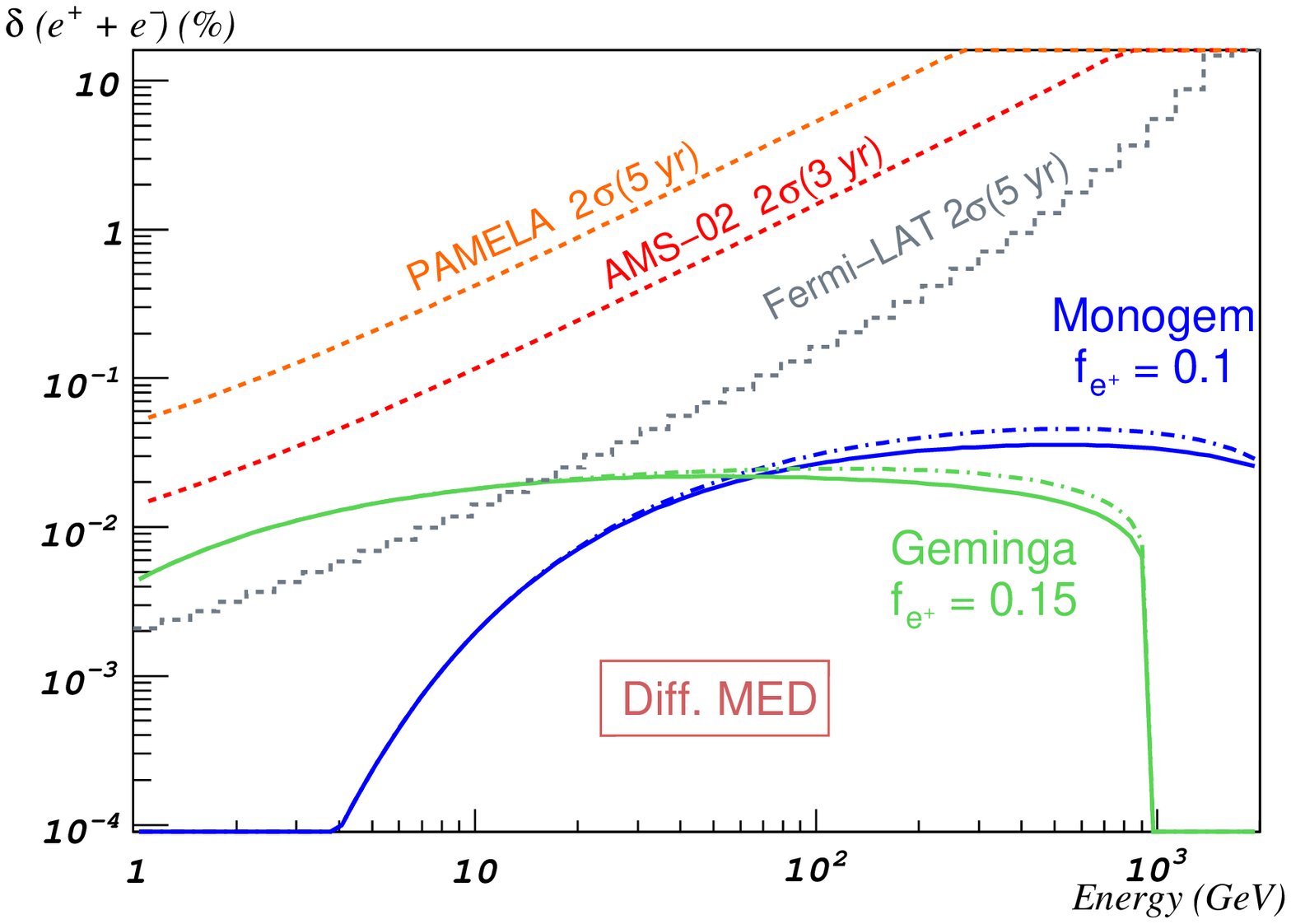}
      \end{minipage}
        \caption{\it Left: Anisotropy in $e^{\pm}$ flux for sum of Geminga and Monogem contributions
	             for MED diffusion. Right: Geminga and Monogem with pulsars background (continued line)
		     and without (dashed line) for MED diffusion with sensitivity expected for PAMELA, AMS-02 and Fermi-LAT.}
        \label{anis_fond}
     \end{figure}

     In previous sections, Geminga and Monogem were supposed to be sufficiently distant from each other to induce
     an individual dipole. Nevertheless, according particles propagation and respective pulsar galactic coordinates, 
     $Monogem (b,l)=(201.11^0,+08.25^0)$ and $Geminga (b,l)=(195.13^0,+04.27^0)$, one can assume that total measured
     anisotropy could be sum of the two. Indeed, propagation could be helpful for anisotropy detection by merging
     Geminga and Monogem electrons. Taking MED propagation set, sum of the two pulsars anisotropy, in figure 
     {\bf \ref{anis_fond}} (left), gives a too weak anisotropy increasing to be considered like a enhancement. 
     
     An another hypothesis which can be discussed is the pulsar background. Faraway pulsars (section {\bf \ref{fond_elec}})
     can be added to the standard electron-positron background $N_e^{tot}$ in the equation {\bf \ref{anis_dep}}.
     Looking at pulsars continuum, propagation must be sufficiently diffuse to produce an isotropic contribution
     overlapping Monogem and Geminga contributions. This new anisotropy is estimated for Geminga and Monogem in figure
     {\bf \ref{anis_fond}} (right) with continued lines, compared to the case without pulsars background in dashed line.
     The biggest effect is expected at high energy where anisotropy would be lower and unreachable with detector
     sensitivity. Therefore, pulsar background should not affect anisotropy measurement. An other aspect must be
     discussed about anisotropy, which is its own standard fluctuation. Indeed, geomagnetic field must produce some
     effects in anisotropy map inducing some natural anisotropy which must be understood before any anisotropy claim.


  \section*{Conclusion}
    
     For almost four years, positron fraction is studied with new generation experiment, PAMELA had confirmed the
     increase of positrons above 10 GeV which can indicate a primary $e^+$ source like pulsars, and Fermi-LAT has
     observed electrons flux with great accuracy and it seems to be above standard background. Next years, AMS-02
     will provide a positron fraction at higher energy with better accuracy than PAMELA, and Fermi-LAT will extend
     his energy range from GeV to TeV. In this article, we considered that positron excess is the result of electron-positron
     pairs production in nearby pulsars, affected by different propagation scenarios. AMS-02 will the only experiment
     able to investigate positrons at high energy, and therefore constraint more efficiently $e^+$ sources. The closest
     pulsars can contribute to positron fraction and propagation should modify contributions, in two extremes ways:
     propagation scenario which maximizes electrons flux (MAX) should allow each pulsar to be noticed, and the other
     hand the one minimizing (MIN) the flux will smooth all contributions giving a continuum flux. The MAX scenario is
     preferred from experimental point of view, and it is in agreement with pulsars models implying low pair production
     parameters, around $\%$. In this configuration, AMS-02 will be the only experiment able to confirm pulsars implication.
     The others  propagation parameters sets will not allow to distinguish each pulsar, and for the one minimizing positron
     flux (MIN), pulsars should not be able to produce the whole positron contribution. Without direct detection of pulsars,
     anisotropy studies in the positron and positron-electron flux will be powerful test to establish pulsar production.
     Geminga and Monogem should induce enough $e^{\pm}$ anisotropy to be detected in five years with 2$\sigma$ significance
     by Fermi-LAT, and AMS-02 will be able to detect an $e^+$ anisotropy at the same level after 3 years if Geminga
     and Monogem are the only contribution to positron excess. Others candidates, like supernova remnant (SNR) or
     Dark Matter, can participate to leptons flux, and  AMS-02 will be a helpful detector studying from leptons to nuclei
     constraining each contributions.
     

\end{document}